\documentclass[fleqn,10pt]{SelfArx} 

\usepackage[english]{babel} 
\usepackage[final]{pdfpages}
\usepackage{eurosym}

\usepackage[T1]{fontenc}
\usepackage[utf8]{inputenc}
\usepackage{authblk}

\usepackage{graphicx,epsfig}

\usepackage{times}
\usepackage{graphics,dcolumn,bm,fleqn,float}
\usepackage{amssymb,amsmath,multirow,rotate,color}
\usepackage{comment} 
\usepackage{cite} 

\definecolor{color1}{RGB}{0,0,90} 
\definecolor{color2}{RGB}{0,20,20} 

\usepackage{hyperref} 
\hypersetup{hidelinks,colorlinks,breaklinks=true,urlcolor=color2,citecolor=color1,linkcolor=color1,bookmarksopen=false,pdftitle={Title},pdfauthor={Author}}

\JournalInfo{~} 
\Archive{} 

\PaperTitle{Resource heterogeneity leads to unjust effort distribution in climate change mitigation} 

\Authors{\normalsize Julian Vicens\textsuperscript{a,b,c}, \normalsize Nereida Bueno-Guerra\textsuperscript{d}, \normalsize Mario Guti\'errez-Roig\textsuperscript{b}, \normalsize Carlos Gracia-L\'azaro\textsuperscript{e}, \normalsize Jes\'us G\'omez-Garde\~nes\textsuperscript{e,f}, \normalsize Josep Perell\'o\textsuperscript{b,c}, \normalsize Angel S\'anchez\textsuperscript{e,g,h,i}, \normalsize Yamir Moreno\textsuperscript{e,i,j,k}, and Jordi Duch\textsuperscript{a,l}}

\affiliation{\textsuperscript{a}\textit{Departament d'Enginyeria Inform\`atica i Matem\`atiques, Universitat Rovira i Virgili, 43007, Tarragona, Spain}}
\affiliation{\textsuperscript{b}\textit{Departament de F\'isica de la Mat\`eria Condensada, Universitat de Barcelona, 08028 Barcelona, Spain}}
\affiliation{\textsuperscript{c}\textit{Universitat de Barcelona Institute of Complex Systems UBICS, 08028 Barcelona, Spain}}
\affiliation{\textsuperscript{d}\textit{Psychology Department, Universidad Pontificia de Comillas, 28049, Madrid, Spain}}
\affiliation{\textsuperscript{e}\textit{Institute for Biocomputation and Physics of Complex Systems (BIFI), University of Zaragoza, 50018 Zaragoza, Spain}}
\affiliation{\textsuperscript{f}\textit{Department of Condensed Matter Physics, University of Zaragoza, 50009 Zaragoza, Spain}}
\affiliation{\textsuperscript{g}\textit{Unidad Mixta Interdisciplinar de Comportamiento y Complejidad Social (UMICCS) UC3M-UV-UZ, Spain}}
\affiliation{\textsuperscript{h}\textit{Grupo Interdisciplinar de Sistemas Complejos (GISC), Unidad de Matem\'atica, Modelizaci\'on y Ciencia Computacional, Universidad Carlos III de Madrid, 28911 Legan\'es, Spain}}
\affiliation{\textsuperscript{i}\textit{Institute UC3M-BS of Financial Big Data, Universidad Carlos III de Madrid, 28903 Getafe, Spain}}
\affiliation{\textsuperscript{j}\textit{Department of Theoretical Physics, University of Zaragoza, 50009 Zaragoza, Spain}}
\affiliation{\textsuperscript{k}\textit{Institute for Scientific Interchange, ISI Foundation, Turin, Italy}}
\affiliation{\textsuperscript{l}\textit{Northwestern Institute on Complex Systems, Northwestern University, 60208, Evanston, USA}}

\newcommand{\keywordname}{Keywords} 

\Abstract{Fighting against climate change is a global challenge shared by nations with heterogeneous economical resources and individuals with diverse propensity for cooperation. However, we lack a clear understanding of the role of key factors such as inequality of means when diverse agents interact together towards a common goal. Here, we report the results of a collective-risk dilemma experiment in which groups of subjects were initially given either equal or unequal endowments. We found that although the collective goal was always achieved regardless of the initial capital distribution, the effort distribution was highly inequitable. Specifically, participants with fewer resources contributed significantly more to the public goods than the richer $-$sometimes twice as much. An unsupervised learning algorithm clustered the subjects according to their individual behavior. We found that the poorest participants congregated within the two ``generous clusters'' whereas the richest were mostly classified into a ``greedy cluster''. Our findings suggest that future policies would benefit both from reinforcing climate justice actions addressed to most vulnerable people and educating fairness instead of focusing on understanding of generic or global climate consequences, as the latter has not proven to drive equitable contributions.}

\begin{document}
\flushbottom 

\maketitle 

\thispagestyle{empty} 

Mitigating anthropogenic climate change \cite{IPCC5} is a complex problem involving many heterogeneous actors with different agendas. In particular, at a global level, some countries are rich and others are poor \cite{azar:96}, and the latter may lack the financial resources to cope with climate change. On the other hand, poor countries will be the most affected by climate change  \cite{Dennig2015,Mendelsohn2006}, while richer countries generate most of the CO$_2$ and cause the biggest impact on climate \cite{Raupach2014}. In so far as climate change mitigation requires a collective action \cite{olson:71}, it is not clear what are the effects of the inherent diversity of the agents involved in it \cite{ruttan:06}. There is the risk that the poor exploits the rich, i.e., that the largest beneficiaries of a common goods bear a disproportionately large burden in its production \cite{olson:71}. Conversely, the poor also have an incentive to contribute since they are risking as much as the rich in the event of a catastrophic development. On the other hand, inequality between actors is a widespread situation not only related to climate change mitigation, but also observed at several socioeconomic levels, e.g., within a given urban and shared space. For instance, increasing green spaces and similar actions that improve urban environmental health are part of the recipes within the Paris Agreement (COP21) \cite{cop21,rhodes:16}. However, these actions impact differently more and less affluent citizens and, in fact, such actions have also been questioned from the perspective of environmental justice \cite{wolch:14}. This is the reason for the increasing interest in linking climate justice with socio-economic inequalities and in incorporating some behavioral aspects in city policy \cite{bulkeley:14}. Therefore, understanding how to deal with climate change mitigation in an economically diverse world and in an environmentally fair manner is of special interest. This has become more timely and pressing with the withdrawal \cite{trump} of the second largest CO$_2$ emitter \cite{edgar} from COP21. 

Resource heterogeneity and environmental justice can be very suitably framed within the collective-risk dilemma experimental setup introduced by Milinski {\em et al.} a decade ago

\onecolumn

\noindent \cite{Milinski2008}. In this framework, to be described in detail below, groups of people have to reach a common goal by making small contributions from an initial endowment. If the goal is reached, every subject keeps the part of the money she did not contribute. If not, a catastrophe occurs with certain probability, and all participants lose all the money they kept. While many experimental and theoretical studies have considered different aspects of climate change within this framework \cite{
Milinski2011,Tavoni2011,abou:12,
santos:12,burton:13,hilbe:13,Jacquet2013,abou:14,
freytag:14,vasconcelos:14,dannenberg:15,bynum:16,hagel:16,milinski:16,milinski:17}, the issue of heterogeneity has only been considered in two experiments. Thus, Tavoni et al. \cite{Tavoni2011}. included inequality by separating the participants into two groups with different starting endowments, and found that the common goal was less likely to be reached. However, when they allowed participants to communicate their intentions, the probability of reaching the target goal increased again, similarly to  conditional cooperation in public goods games \cite{Fischbacher2001}. On the other hand, Milinski {\em et al.} \cite{Milinski2011} observed that when they included rich and poor subjects, rich ones substituted for missing contributions by the poor provided intermediate climate targets.
were established. However, despite this increase in the contributions of the rich, the final target was reached less often than the intermediate target.  

Here, we largely extend the knowledge on the effects of resource heterogeneity in collective dilemma in two main directions. First, we include broader capital distributions, thus representing more closely the diversity in resource availability among the members of a given collective public goods dilemma, e.g., between different countries worldwide or inhabitants in a given urban context. Second, and most importantly, we go beyond aggregate results to analyze the behavior of individuals by means of agnostic classification tools that allow us to identify differences between the behavior of subjects with the same resources. We complement this with a questionnaire probing into the subjects' knowledge of the climate change crisis and the influence of such knowledge in their actions. Therefore, we provide a much more complete picture of the influence of inequality that encompasses both the global (reaching the collective goal) and the local (how different individuals behave under different circumstances) visions of the problem. Such a two-level approach is the best option in order to identify how agents react to resource heterogeneity and what actions must be taken to promote environmental justice. As we show below, our findings allow to hint directions for policy measures targeted to specific collectives.

\section*{The collective risk dilemma game}

\subsection*{Game definition} The original collective-risk dilemma \cite{Milinski2008} introduced groups of six people where each person receives an initial capital (40 \euro{}). The common goal of the group is to collect 120 \euro{} that will be invested in mitigating climate change (by publishing an ad in a national newspaper). The game consists of 10 rounds, and at every round each subject decides how much she contributes to the common fund (0, 2, or 4 \euro{}). If the goal is reached at (or before) the end of the game, all participants keep the money that they have not contributed. Otherwise, the participants only keep the remaining money with a probability which in \cite{Milinski2008} was 90\%, 50\%, or 10\% (equivalently, a climatic catastrophe occurred with probability 10\%, 50\%, or 90\%). In addition, in this case no money goes to climate change mitigation. The main result was that most groups did not reach the goal, and even in the worst case scenario (catastrophe probability 90\%) only about half of the groups avoided climate change. We chose this worst case scenario as our baseline treatment, and we carried out experiments with the homogeneous distribution for comparison with the heterogeneous one.

To introduce inequality, we randomly assigned one of six different starting capitals (20, 30, 40, 40, 50, and 60 \euro{} to each participant. In half of the games the participants could invest 0, 2, or 4 \euro{} per round as in the above setup, while in the other half of the games we allowed them more flexible choices, namely 0, 1, 2, 3, and 4 \euro{} per round (see SI Sec.\ 2). In all cases we informed the participants that in case they reached the goal of 120 \euro{}, the so collected money would be used for a reforestation action by planting trees in a nearby park with the help of an NGO organisation \cite{ngo}. Finally, after the experiment, we asked our subjects to answer a questionnaire (see SI Sec.\ 8) to have an individual assessment of climate change awareness and predisposition to collaborate in common actions that could be further correlated with the individual's contributions.

\subsection*{Equilibria and fair distribution}

In our heterogeneous version of the collective-risk dilemma there are very many Nash equilibria \cite{nash}, which makes claiming that one or other behavior should be expected very difficult. Indeed, in the homogeneous case, the number of equilibria can be refined by imposing symmetry (meaning that all subjects, being equal, should choose the same contribution). Then, two equilibria are left, with subjects either contributing nothing or contributing exactly 20 \euro{}. Of course, these equilibria refers only to accumulated contribution along the game; considering the different sequences of investments in the 10 rounds recovers the multiplicity of equilibria and we will not consider them. In our heterogeneous setup we have 5 types of players (there are two endowed with 40 \euro{}) and the symmetry refinement no longer holds. Subjects contributing nothing is still a Nash equilibrium that leads to expected gains of 10\% of every subject's endowment. It is then easy to show that any combination of individual total investments that adds up to exactly 120 \euro{} such that every player makes more money than in the "contribute nothing" case is also an equilibrium. Therefore, there is not a clear theoretical prediction about what should happen in our heterogeneous version of the game. In this situation, we decided to take as a reference for discussion what we call the "fair equilibrium", which is an analogous to the symmetric case. In this equilibrium, contributions correspond to equal common efforts, assuming 12 \euro{} total contribution per round. Then, at the individual level, we considered fair a 50\% contribution of every subject's initial endowment (i.e., 1 \euro{} per round for participants starting with 20 \euro{}, 1.5 \euro{} per round for participants starting with 30 \euro{}, and so on; note that fractions of \euro{} are not possible but participants can alternate, e.g., between contributions of 1 and 2 to achieve it). This equilibrium reduces to the symmetric equilibrium when we go back to the homogeneous situation, which is yet another reason to use it as our reference point. 

\section*{Results}

\subsection*{Collective climate action}

\begin{figure}[!tb]
\begin{center}
\includegraphics[width=0.6\columnwidth]{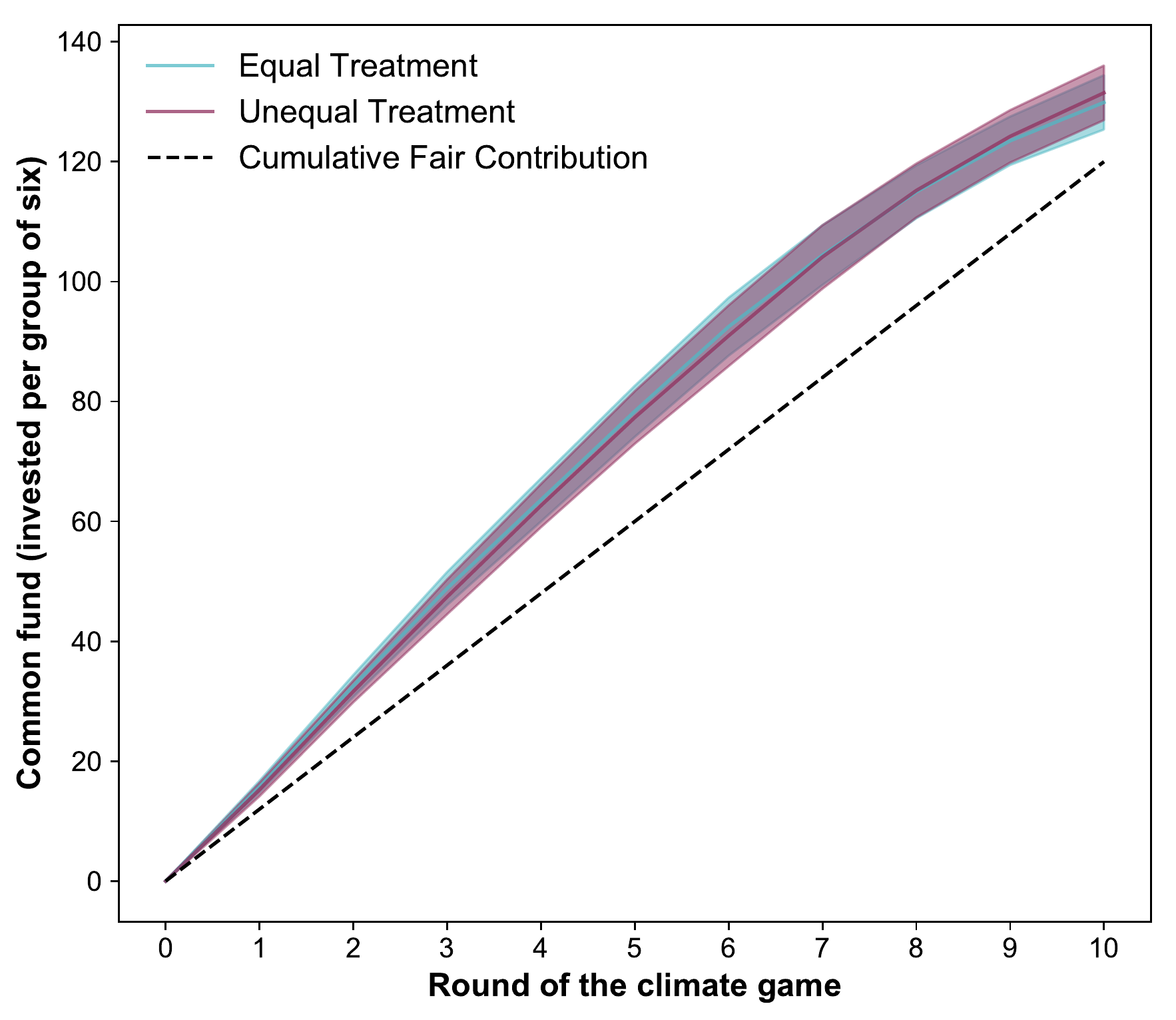} \caption{Average (CI=95\%) contribution to common fund per round in the Climate Game for Equal treatment (all players are endowed with 40 \euro{}) and Unequal treatment (endowments are 20, 30, 40, 40, 50, and 60 \euro{}). Both treatments show an accumulated contribution over the game evolution above the fair contribution per round.}
\label{fig:common_funds_evolution}
\end{center}
\end{figure}

In all games played in our experiment, the participants reached the goal irrespective of the initial endowments being homogeneous or heterogeneous. In the former case, our result has to be compared with only 50\% success rate for the groups in \cite{Milinski2008}. An increase in groups reaching the target has been observed in other similar studies carried out later \cite{Milinski2011}. The evolution of the games also differ from previous experiments (see Fig. \ref{fig:common_funds_evolution}): In all the treatments in our experiment, the sum of money accumulated at the end of each round is always above the fair contribution (12 \euro{} per round), that is, participants contribute much more in the initial rounds, making the group reach the goal faster, and then they stop contributing at the end once the goal has been secured (see SI Fig. S2 and Fig. S5). In contrast, the original results in \cite{Milinski2008} showed contributions below the fair one until the end of the game, and those groups that reached the goal did it by increasing their contributions in the last rounds. 

\subsection*{Effect of unequal capital distribution}

\begin{figure}[!htb]
\begin{center}
\includegraphics[width=0.99\columnwidth]{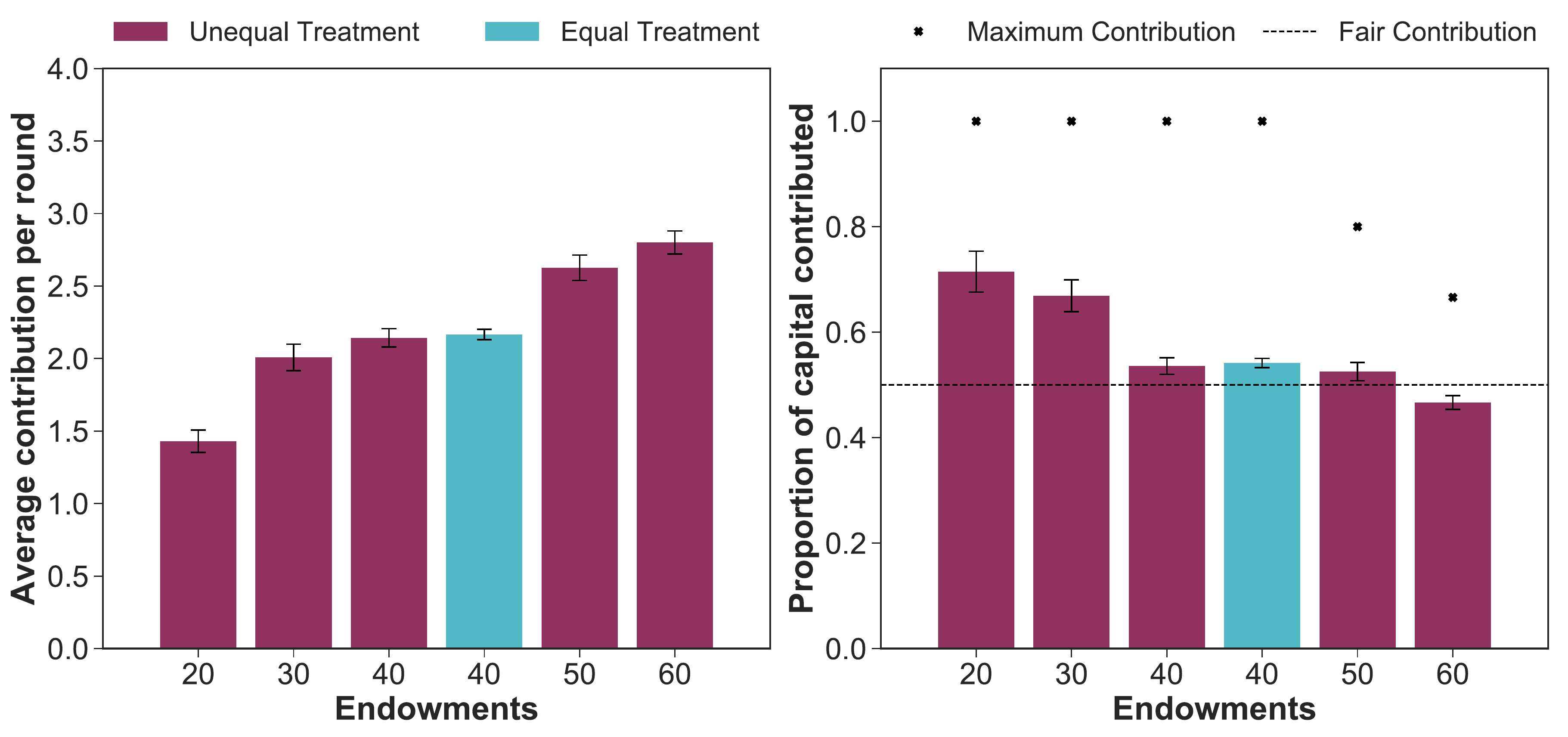} 
\caption{(Left) Average capital contributed and standard error of the mean by players according to their assigned initial capital. Note that players starting with 20 \euro{} and 30 \euro{} can only reach a maximum average contribution per round of 2 \euro{} and 3 \euro{} respectively. (Right) Average proportion of capital contributed and standard error of the mean in the Equal and Unequal treatments. Dotted line represents the fair contribution, which we have defined as contributing 50\% of the initial capital. The effort to contribute is different depending on the endowments, so black crosses represent the maximum investment that each group can reach. Subjects with endowments of 50 and 60 \euro{} always keep a proportion of capital as savings even if they contribute the maximum amount of 4\euro{} per round.}
\label{fig:average_contribution}
\end{center}
\end{figure}

Figure \ref{fig:average_contribution}  presents the average amount of capital contributed as a function of the initial capital of the participants. We observe that, in terms of absolute contribution, the subjects with high endowments, 50 and 60 \euro{}, are the ones that contribute (mean$\pm$sem) the most, $2.6\pm0.08$ \euro{} and $2.8\pm0.08$ \euro{} per round respectively. Subjects with low-endowments, 20 and 30 \euro{}, contribute the least,$1.43\pm0.08$ \euro{} and $2\pm0.09$ \euro{} respectively. However, this comparison is not a proper one, since the initial capital of the poorest players only allows them to contribute a maximum of 2 \euro{} per round (earning 0 at the end). Therefore, the comparison makes more sense in terms of the percentage of capital contributed relative to their total capital, which in turn allows to discuss the results using the fair distribution as reference. Strikingly, we observe that the most affluent (endowments of 60 \euro{}) are the ones that contribute proportionally less, with around 46.6\% of their initial capital, while the poorest (starting with 20 \euro{}) contribute around 71.4\% of their initial capital which shows their vulnerability when facing the collective risk dilemma. Figure \ref{fig:average_contribution} shows very clearly the stark contrast between the two visions. To put this result further in context, we notice that the maximum contribution from participants with a starting capital of 20 \euro{} (2 \euro{} per round) implies an effort of 2 times the fair contribution, whereas for participants with a starting capital of 60 \euro{}, the effort of contributing 2 \euro{} per round is 0.66 times the fair contribution. Therefore, in that case, contributing 2 \euro{} have unequal impacts in the endowments of subjects: poor people, contributing 1.43 \euro{} on average, make a larger effort than that of the rich contributing 2.8 \euro{}, which is even below their fair share of the threshold. 

\begin{figure*}[!tb]
\begin{center}
\includegraphics[width=0.8\textwidth]{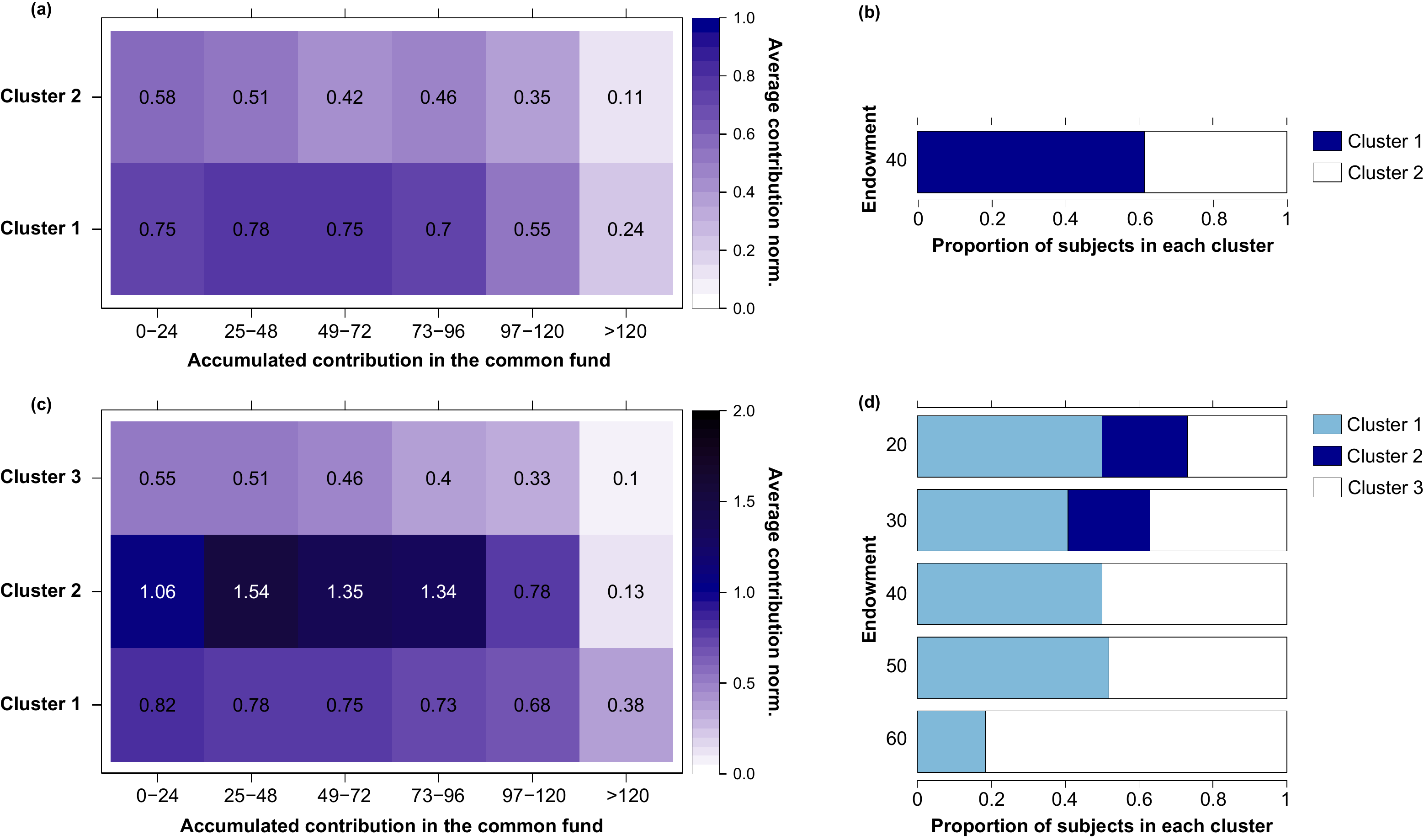}
\caption{Equal treatment. (a) Behavioral patterns based on average contribution during the evolution of the game. The value in each cell represents the average contribution normalized by the initial capital per round (i.e: 2 for participants starting with 20 \euro, 3 for participants starting with 30 \euro, and so on; 0.5 is the fair contribution) in a given stage of the game. (b) Cluster 1 is formed by generous subjects (61\%) with average contribution above the fair while cluster 2 is formed by subjects (39\%) that contribute around and below the fair contribution. Unequal treatment. (c) Behavioral patterns based on average contribution during the evolution of the game. The value in each cell represents the average contribution normalized by the inital capital per round in a given stage of the game. Cluster 1 is formed by generous subjects (43.48\%) with average contribution above fair, Cluster 2 consists of subjects (7.45\%) that contribute very much above fair, and Cluster 3 is formed by subjects (49.07\%) that contribute around and below the fair contribution. (d) Distribution of the different types among the subjects as a function of their initial endowment.}
\label{fig:clustering}
\end{center}
\end{figure*}

\subsection*{Individual behaviors}

Once we have looked at the average evolution of the different groups, we focus on the individual behavior and study whether participants with the same starting capital behave similarly. We characterize the set of decisions taken by each participant with a vector, grouping decisions by the capital on the common fund at the beginning of the round (see SI Sec.\ 3  for a more detailed explanation). This is motivated by the intuition that subjects choose their contributions as a function of their endowment but also taking into account the current situation and whether the goal is closer or farther. In turn, we can monitor how the contributions differ depending on the stage of the game where they are (see SI Sec.\ 3). Next, in order to detect, identify and characterize different types of behavioral patterns, we use an unsupervised learning algorithm, namely Ward's hierarchical clustering method \cite{Ward1963, Murtagh2014} with squared Euclidean distances. Additionally, we used a consensus clustering to look for the optimal subdivision of our data into groups as well as for the robustness of each group (see SI Sec.\ 4). This allowed us to find the groups that better fit the collected data as well as a much more stable solution \cite{Monti2003}. 

In Fig.\ref{fig:clustering}(a) we present the results of the clustering of the participants of the control group, with equal capital distribution. In this scenario the optimal number of groups identified is two. Looking at the results of the clustering we immediately identify two different types of behaviors, a group of generous participants (cluster 1) that contribute above the fair contribution, and a group of more greedy participants that contribute around the fair contribution at the beginning of the game and decrease their contribution as they approach the end of the game (but before reaching the goal). 

The results for the unequal treatment show that the optimal division of the participants is into three groups, being clusters 1 and 3 those gathering the majority of subjects (92.54\%). Again, in cluster 1 the subjects contribute on average more than the fair contribution, whereas in cluster 3 the average contributions are around the fair value at the beginning, but as the game approaches the end they decrease below the fair amount. A minority group of subjects (7.45\%) is hyper-generous and contributes far beyond what we are considering fair. 

In this latter framework, subjects have different initial endowments. Thus, it is interesting to check how these subjects are distributed in each of the three groups based on their relative contribution. Fig.\ref{fig:clustering}(d) shows that subjects with fewer resources (20-30 \euro{}) than the rest are concentrated in the generous clusters (1 and 2). In fact, the cluster 2 (the hyper-generous group) is formed exclusively by subjects with low endowments. On the other hand, the third cluster is mainly composed by subjects with mid and high endowments. This means that the majority of low endowed participants, 73.07\% (20 \euro{}) and 62.96\% (30 \euro{}) contributed above the fair threshold, different from the subjects with high endowments, where only 51.85\% (50 \euro{}) and 18.51\% (60 \euro{}) contributed a fair amount. Interestingly, the comparison of Figs.\ \ref{fig:clustering}(b) and (d) shows that subjects with mid endowments (40 - 50 \euro{}) distributed among the two clusters not very differently from the equal treatment. Therefore, the richest participants were those who diverged from that distribution. 

\subsection*{Effect of awareness about climate change}
Since all groups reached the goal, one could be tempted to conclude that the difference with \cite{Milinski2008}'s results is that awareness about climate change has spread among the general population within the decade that elapsed since then. This might be certainly true in a broad sense. However, and very surprisingly, the majority of our sample (N=294, 91.3\%) failed basic questions about climate change concepts (see SI Sec.\ 8), in a questionnaire that included basic questions about the greenhouse effect, carbon footprint, or the Kyoto Protocol. We stress that this is so even if the experiment was done the week following the COP21 summit in Paris, which had led us to expect much more familiarity of our subjects with climate change. This result allows us to exclude that more literacy and more public outreach efforts on climate change are the reasons for the participants reaching the goal in all cases. 

\subsection*{Effect of socio-demographics and beliefs} To begin with, no significant differences were observed in terms of age or gender of the participants, i.e., the distribution in generous (including hyper-generous) and selfish subject is independent of those variables. Next, we divided the sample into graduates and non-graduates and explored whether a more educated level predicted greater contributions to the common pool. A GLMM \cite{Baayen2008} showed that educational level was a factor that significantly affected the average contribution (GLMM, $\chi^2$=3.811, df=1, p=.006). Post hoc tests revealed that subjects with lower education level were predicted to make higher contributions in equal conditions (linear regression, F(1,156)=7.219, p<.05). On the other hand, a third part of the sample (N=112, 34.8\%) expected to arrive to the common goal before starting the game. Harboring this previous expectation did not have an influence over their average contribution ($\chi^2$= 6.005, df=3, p=.111). 

An important point is that, as we saw above, the average contribution depended on the individual profile in the three categories detected (cluster 1, generous; cluster 2, hyper generous; cluster 3, greedy) since subjects did not seem to react to the money that was being collected. This is consistent with the fact that the majority of the sample (around 87\%) claimed that their contributions did not have to match the co-participant's contribution, showing absence of conditional cooperation \cite{Fischbacher2001}, at least in climate-related collective goals. Also, the idea of fair contributions does not fit the real contributions to the common fund: more than half of the participants (N=214, 66\%) agreed with the statement ``Contributions should be proportional to the initial capital so that players with more capital should contribute more to the pool''. In contrast, when analyzing their responses in unequal conditions, those who firmly agreed with proportionality contributed less than those who did not strongly support that (Z= -2.653, p<.05), especially when their initial endowment was high. For example, the participants with a starting capital of 60 \euro{} that adhered to "proportionate contributions" contributed an average of 2.6 \euro{}/per round, whereas those with the same initial endowment who did not firmly claim that contributed 3.05 \euro{} per round.

\subsection*{Effect of generosity on emergence of inequality} Finally, we have measured the impact of the observed higher levels of cooperation among the poorest participants on the capital distribution. In the homogeneous treatment, where everybody starts with exactly the same endowments, the Gini Coefficient increases from 0 to 0.1806 at the end of the experiment. Similarly, in the heterogeneous treatment the Gini coefficient also increases around 0.18, however in this case it is much more dramatic since it starts with an inequality accounting for 0.1817 and ends with 0.3483. It is important to keep in mind in this respect that a given behavior consisting in contributing the fair amount would not change the initial Gini coefficient and the levels of inequality would thus not increase (see SI Sec.\ 5).

\section*{Discussion}

Our experimental results allow us to conclude that heterogeneity may lead to unexpected problems in climate change mitigation, mostly related to environmental justice. We observed that all groups reached the goal, which, compared to 50\% in \cite{Milinski2008} less than ten years ago, seems quite an improvement. In this respect, we believe that there are several factors than can intervene in this increase, ranging from cultural aspects (subjects were German in \cite{Milinski2008} while ours are Spanish) to different perception of the effects of climate change (like in the case of Northern US, where climate change is perceived as less threatening \cite{perception}). Interestingly, knowledge about climate change facts did not play a relevant role in the average contributions to the common fund. This might indicate that policies mainly addressed to increase climate change awareness might not be the most efficient solution to foster cooperation, and suggests that emphasizing a correct interpretation of the perceived effects might be more useful in this regard. 

Even if all groups avoided the climatic catastrophe, our work shows that other potentially serious issues may arise in the process. A particularly important one is that poor subjects are contributing much more than a fair share of the mitigation, and that the richest ones are contributing less. It is telling that all hyper-generous behavior is observed in the two poorest types of individuals, while a large majority of those endowed with the largest amount behaved selfishly (irrespective of what they claim to believe about fair contributions, as we have seen). It thus appears that, contrary to the expectations of the poor exploiting the rich in a public goods context, here we found the opposite situation. At this point, it is important to note that these experiments were done in a short period of time and therefore they do not inform about long-term behavior. It would be possible that the fact that the poor contributed more for a long time might eventually lead them to stop doing so, thus jeopardizing the collective goal. The decisions we observed in the experiment did not allow us to learn about the behavior of the richest players in that case, i.e., whether or not they would jump in to solve the problem (although the results in \cite{Milinski2011} suggest that this could be the case). 

Finally, it is important to discuss the different behaviors observed in the collectives we have worked with. Education does not help here: less educated-less favored participants contributed more to the collective goal than more educated-more favored ones. This could indicate that there is an underlying cultural assumption of sacrifice of the most disadvantaged people (related to their vulnerability): in a situation where the poorest are the ones who will face the worst consequences, more advantaged participants may feel inclined to contribute less to solving the problem. Particularly alarming is the fact that, in the group of the richest participants, about 80\% behave in a selfish manner. As this is the group that has the largest means to help to mitigate climate change, their fault to contribute may jeopardize the whole society, which calls for specific actions to work with this segment of the population. It thus seems that, aside from our earlier remark about education on climate awareness, a good general education is also not the remedy to avoid this kind of inequality in contribution. Future policies would benefit from public transparent data of national contributions and a systematic action to avoid unbalanced contributions, both at the national and at the individual levels. In addition, making everybody aware that mitigating climate change must be done within climate justice criteria is necessary if we are to avoid exploitation of the poor.

\section*{Methods}

The experiment was conducted following the lab-in-the-field experiment guidelines used in \cite{Poncela-Casasnovas2016a}, which helped us recruiting participants form a general audience, opposed to the typical samples of university undergraduate students. All participants in the experiment signed an informed consent to participate and no association was ever made between their real names and the results, in agreement with the Spanish Law for Personal Data Protection. This procedure was approved by the Ethics Committee of Universidad Carlos III de Madrid. The experiment was conducted in different sessions in the DAU fair in Barcelona during two days (December 12-13, 2015). The total number of games was 54, the number of participants in our experiment was 320 valid subjects (134 women, 41.8\%), adding up to a total of 3240 game decisions collected. If some participant was non-responsive, the experimental platform took over and make the contribution randomly for her; in that case the data was labeled and the subject's decisions were discarded in the analysis, leading to a total number of valid decisions of 3200. The age ranged from 11 to 73 years (mean: 32, SE$\pm$ 13.1). Almost half of the sample was graduated (48\%), whereas the other half was equally distributed between different educational levels (i.e. professional training (16\%), elementary (11\%), middle (11\%) and high school (12\%). Most of the participants were naive to social experiments (N=279, 86\%). Average (SD) earnings were 18.21\euro{} (8.8) (see SI Sec.\ 6) and the average (SD) duration of a game (considering only the time of decision making) was 82.21s (20.98), hence the average round time was 8.22s (see SI Sec.\ 7)

Before starting the game, all the participants were shown a brief tutorial in the tablet in which the experiment was implemented (see SI Sec.\ 9). Researchers present in the experiment reviewed the instructions with them to guarantee they were understanding the basics of the experiment. Participants were reminded that they had to make a decision on each round on how much money they want to contribute to the common goal, but they were not instructed in any particular way nor with any particular goal in mind. 

The subjects participated in groups of six players, each subject was assigned with an initial capital (20 to 60 \euro{}), and the goal of the game was to contribute 120\euro{} on a common fund between all of them. The subjects contributed into the common fund (0 to 4 \euro{}) during 10 rounds, at the end of each round, all players saw the information of how much money has been contributed to the common fund, they also saw the individual contributions of the six players in the previous round and the total amount contributed by each player up to this round (see SI Section Tutorial Fig.S2(f)). If the goal was reached at the end of the game, all the participants kept the money that they had not contributed. The 120 \euro{} collected in the pot will be used once the research is published in organising an event that includes (1) an action against climate change (the action will consist in planting trees in a forest in Barcelona within the Day of the Tree event organised by an NGO\cite{ngo}) and (2) the dissemination of the current results. Otherwise, if the common fund did not reach 120 \euro{} at the end of the game, we didn't take any action against climate change and the participants only kept the remaining money with a probability of 10\%. Once the game finished and if the subjects achieved the goal, they were rewarded with a gift card worth the savings, the capital not contributed.

\section*{Acknowledgements} We acknowledge the participation of 324 anonymous volunteers who made this research possible. We are indebted to the Citizen Science Office promoted by the Direction of Creativity and Innovation from the Institute of Culture of the Barcelona City Council for their help and support for setting up the experiment at the DAU Barcelona Festival at Fabra i Coats. We specially want to thank I Bonhoure, O Mar\'in from Outliers, N Fern\'andez, and P Lorente for all the logistics to make the experiment possible and to O Comas (director of the DAU) for giving us this opportunity. We are grateful to Zhen Wang and Marko Jusup for useful discussions and for sharing some preliminary results of a related study. This work was partially supported by MINECO (Spain) through grants FIS2013-47532-C3-1-P (JD), FIS2016-78904-C3-1-P (JD), FIS2013-47532-C3-2-P (JP, MGR), FIS2016-78904-C3-2-P (JP), FIS2015-71582-C2 (JGG); by MINECO and FEDER funds through grants FIS2014-55867-P (JGG, YM) and FIS2015-64349-P (AS); by Comunidad de Arag\'on (Spain) through FENOL (CGL, JGG, YM); by Generalitat de Catalunya (Spain) through Complexity Lab Barcelona (contract no. 2014 SGR 608, JP and MGR) and through Secretaria d'Universitats i Recerca (contract no. 2013 DI 49, JD, JV); and by the EU through FET Open Project IBSEN (contract no.~662725, AS, YM), and FET-Proactive Project DOLFINS (contract no. 640772, CGL, YM, AS).

\section*{Author contributions}
JP, AS, YM, and JD conceived the original idea for the experiment; 
JV, NBG, CGL, JGG, JP, AS, YM, and JD contributed to the final experimental setup;
JV and JD wrote the software interface for the experiment;
JV, MGR, CGL, JGG, JP, JD carried out the experiment;
JV, NBG, MGR, CGL and JGG analyzed the data;
JV, NBG, MGR, CGL, JGG, JP, AS, YM, and JD discussed the analysis results;
JV, NBG, MGR, CGL, JGG, JP, AS, YM, and JD wrote the paper.

\newpage


\renewcommand{\thefigure}{S\arabic{figure}}
\setcounter{figure}{0} 

\section*{SUPPLEMENTARY INFORMATION}

\section*{Experiment demographics}

\begin{figure}[h]
\begin{center}
\includegraphics[width=0.5\textwidth]{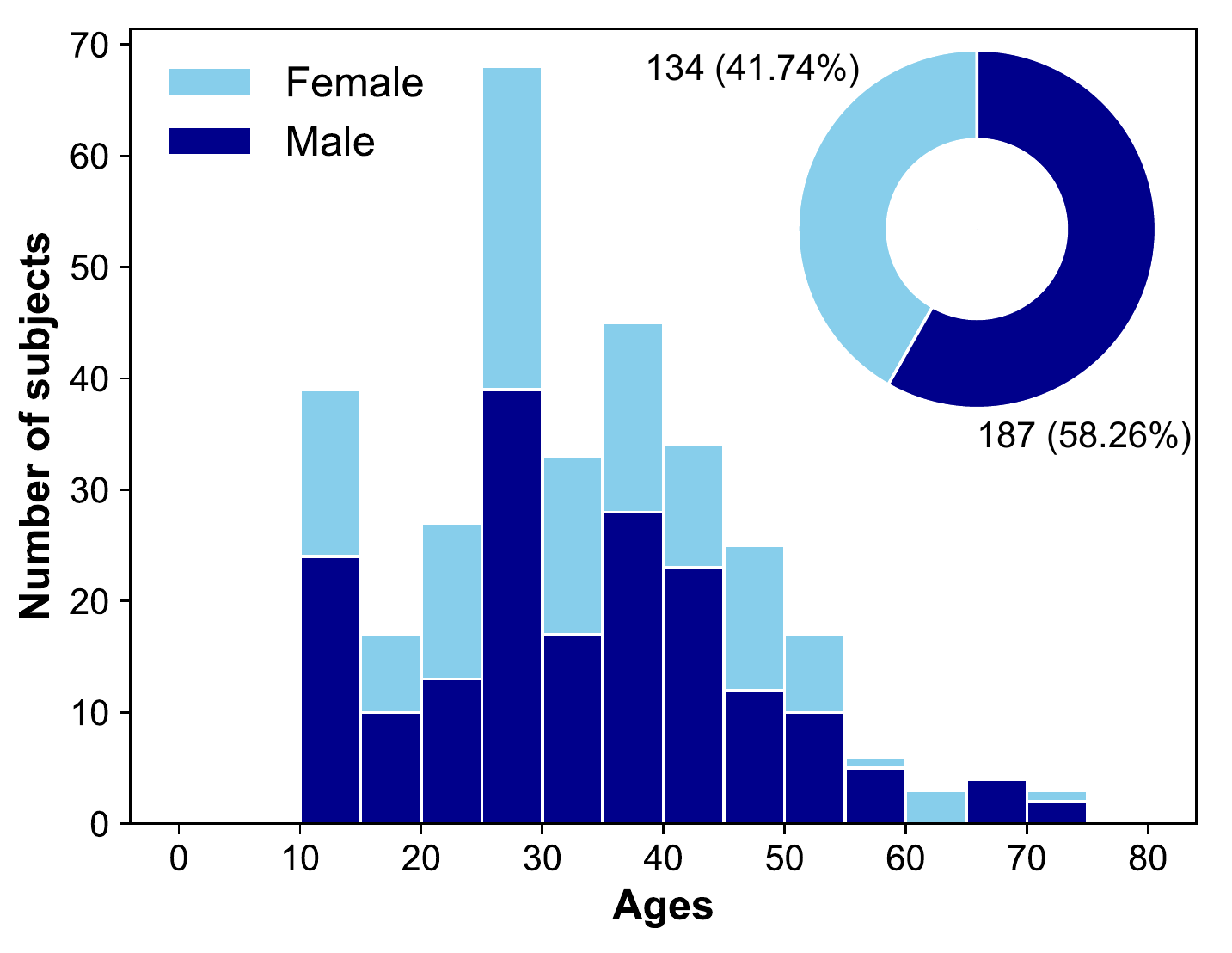} 
\caption{Distribution of subjects in the experiment by age and gender.}
\label{fig:demography}
\end{center}
\end{figure}

The experiment was performed during the games festival (Festival de Jocs de Taula) DAU in Barcelona, in December 2015, over a period of two days. We collected data from 324 subjects who were randomly recruited, of which 320 subjects contributed valid data so 3200 valid decisions. The average age among our participants was 32.28 (SD=13.17), with a proportion of 41.88\% females and 58.12\% males (see Fig. \ref{fig:demography}).


Each participant was given a tablet to play the game on. Before the actual experiment started, the subjects were passed a tutorial (see Sec.\ 9) with the game instructions, to understand the basics and mechanics of the games, the goal and the consequences of their contributions. Also, some of our team members were checking that the participants understood the instructions during the tutorial period (but not during the actual game). Once the participants had understood it, they pressed the button to indicate the system that they were ready to start playing. Once a group of six participants was ready, we started the game.

\subsection*{Effect of gender by contribution}
The average (SD) contribution by gender is: 2.24\euro{} (0.68) in females and 2.13\euro{} (0.72) in males. There are no significant differences in the means by gender (t-test, t: -1.35 p: 0.18). 

\section*{Selection strategy}

\begin{figure}[!h]  
\begin{center}
\includegraphics[width=0.24\textwidth]{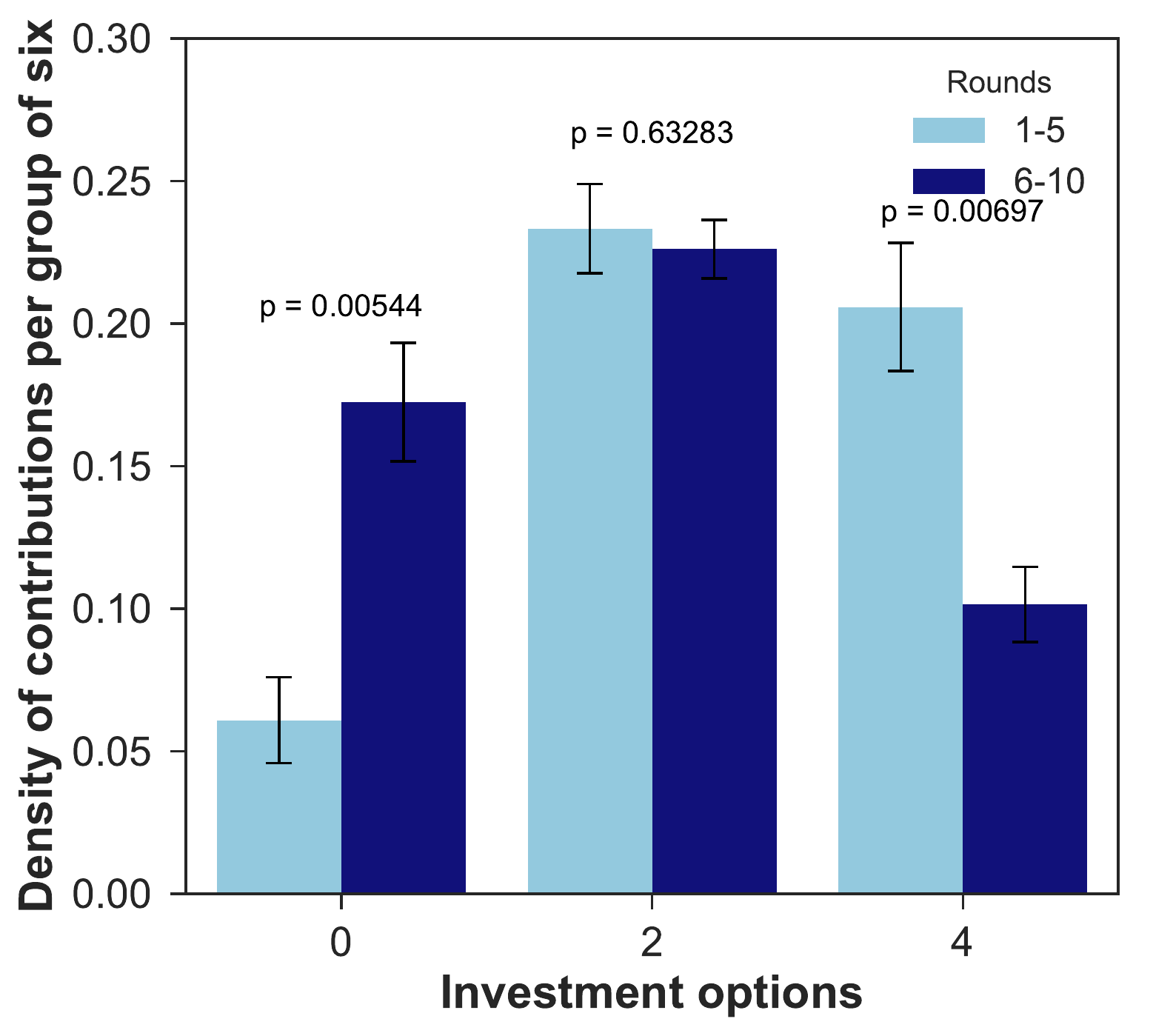}
\includegraphics[width=0.24\textwidth]{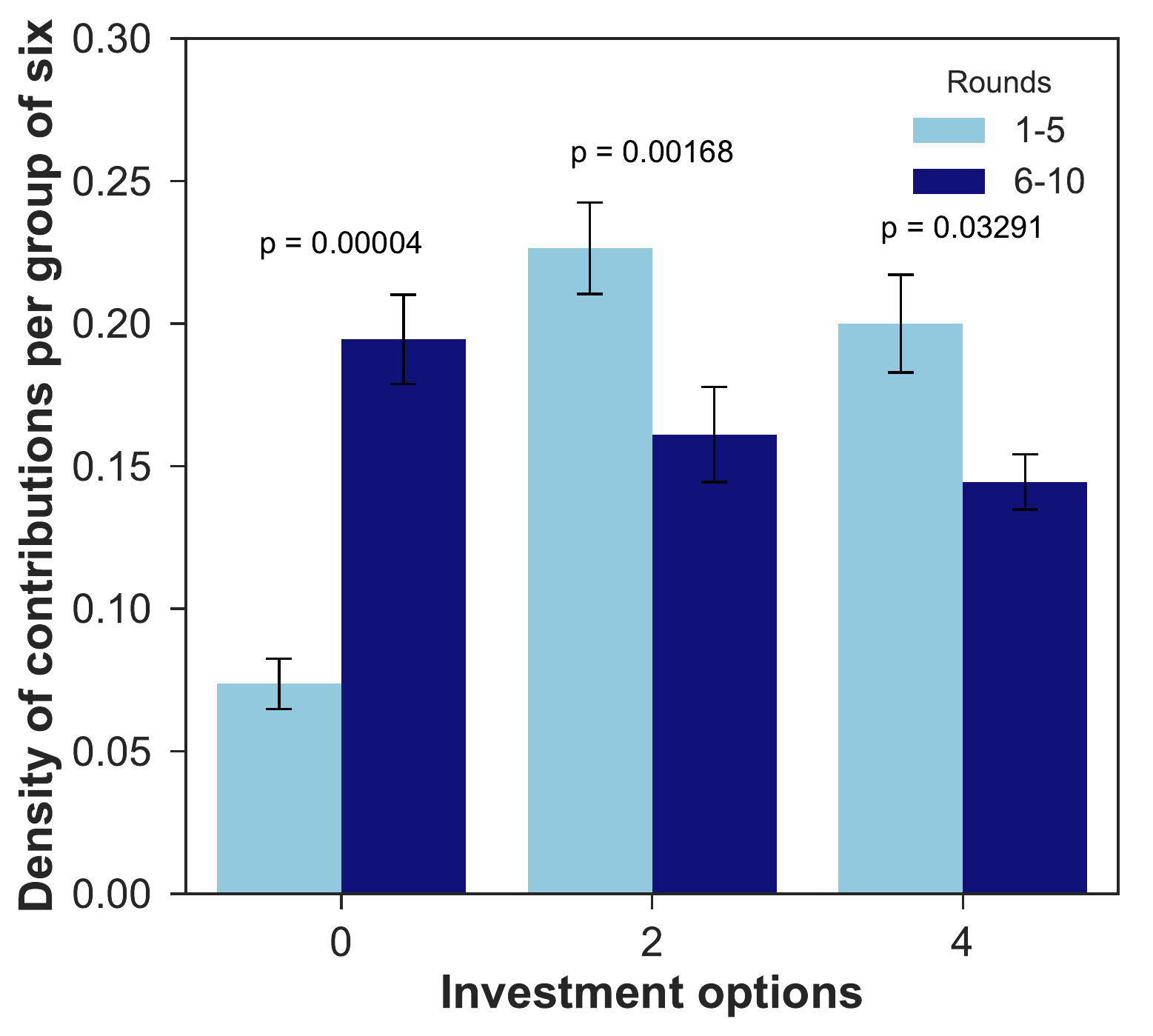}
\includegraphics[width=0.24\textwidth]{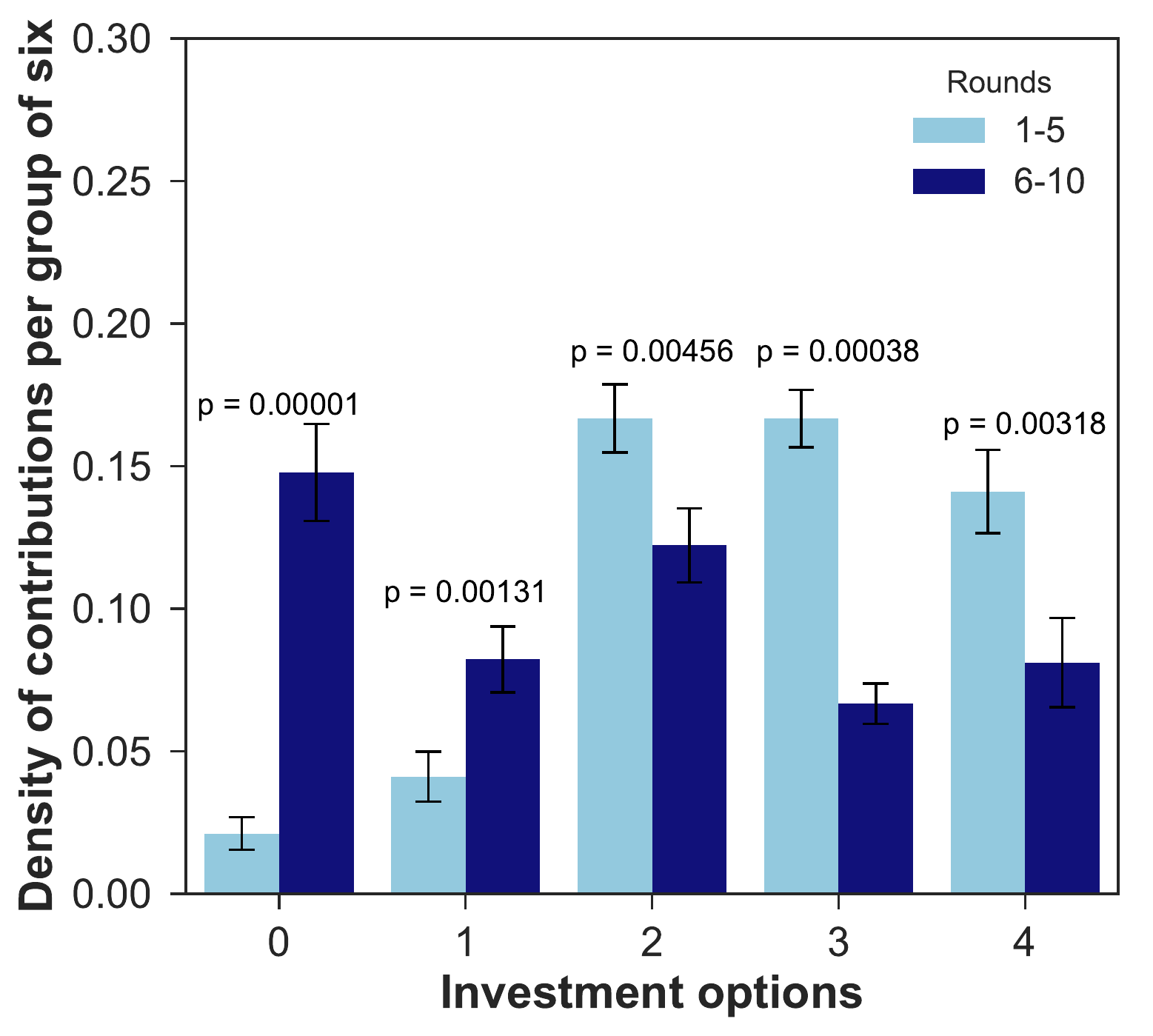}
\includegraphics[width=0.24\textwidth]{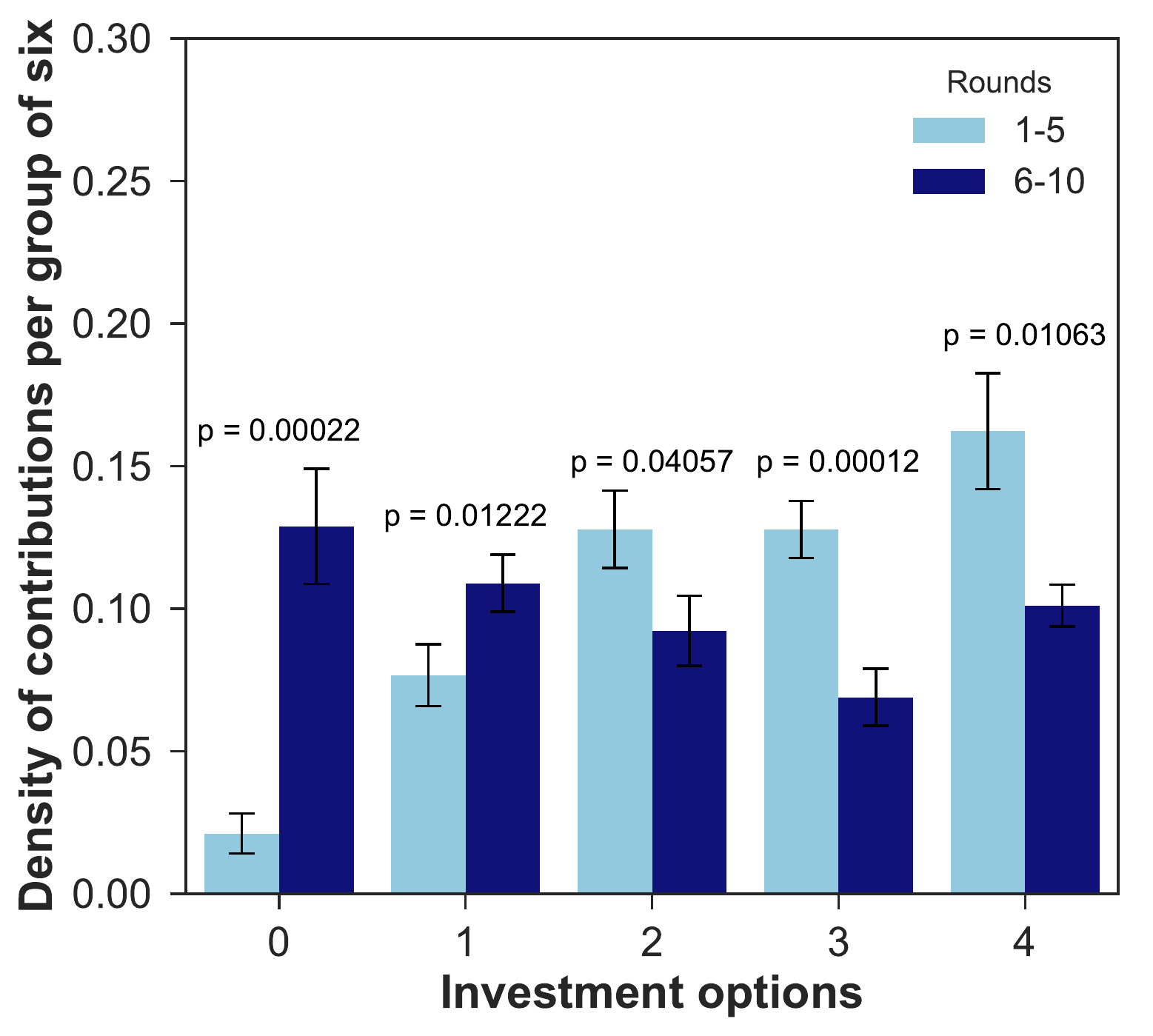}
\end{center}
\caption{Density of investment selections (mean and sem) in the first five rounds and the last five rounds, equal treatment (left column) and unequal treatment (right column).}
\label{fig:density_contribution_by_groups_of_rounds}
\end{figure}

We also studied the effect of including more decision options to the participants, instead of allowing to invest 0, 2 or 4 \euro{} (control setting) the participants could select to invest between 0 and 4 \euro{} (intervention setting). Offering more options to invest allows the existence of more complex investing strategies, since the participants can tune more accurately how much they want to contribute in each round (see Fig. \ref{fig:density_contribution_by_groups_of_rounds}). 


The capital of participants decreased over the game in a different way depending on their endowments. In the unequal treatment, the capital of participants with low endowments decreased faster than the wealthy subjects, partly because the effort to contribute is greater for the poor participants and every single contribution reduced substantially their capital in comparison with the others participants.

\begin{figure}[!hb]  
\begin{center}
\includegraphics[width=0.4\textwidth]{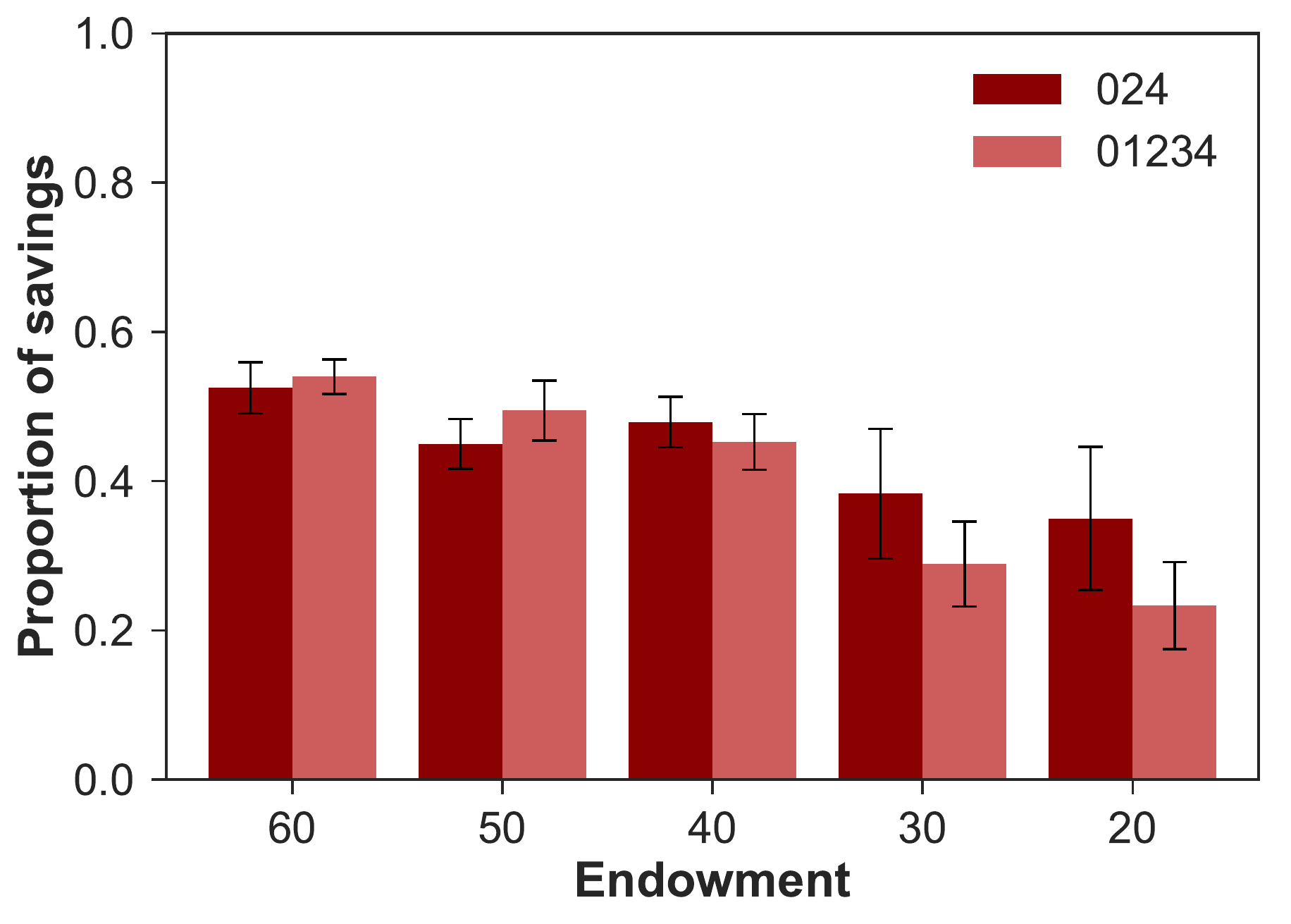}
\includegraphics[width=0.4\textwidth]{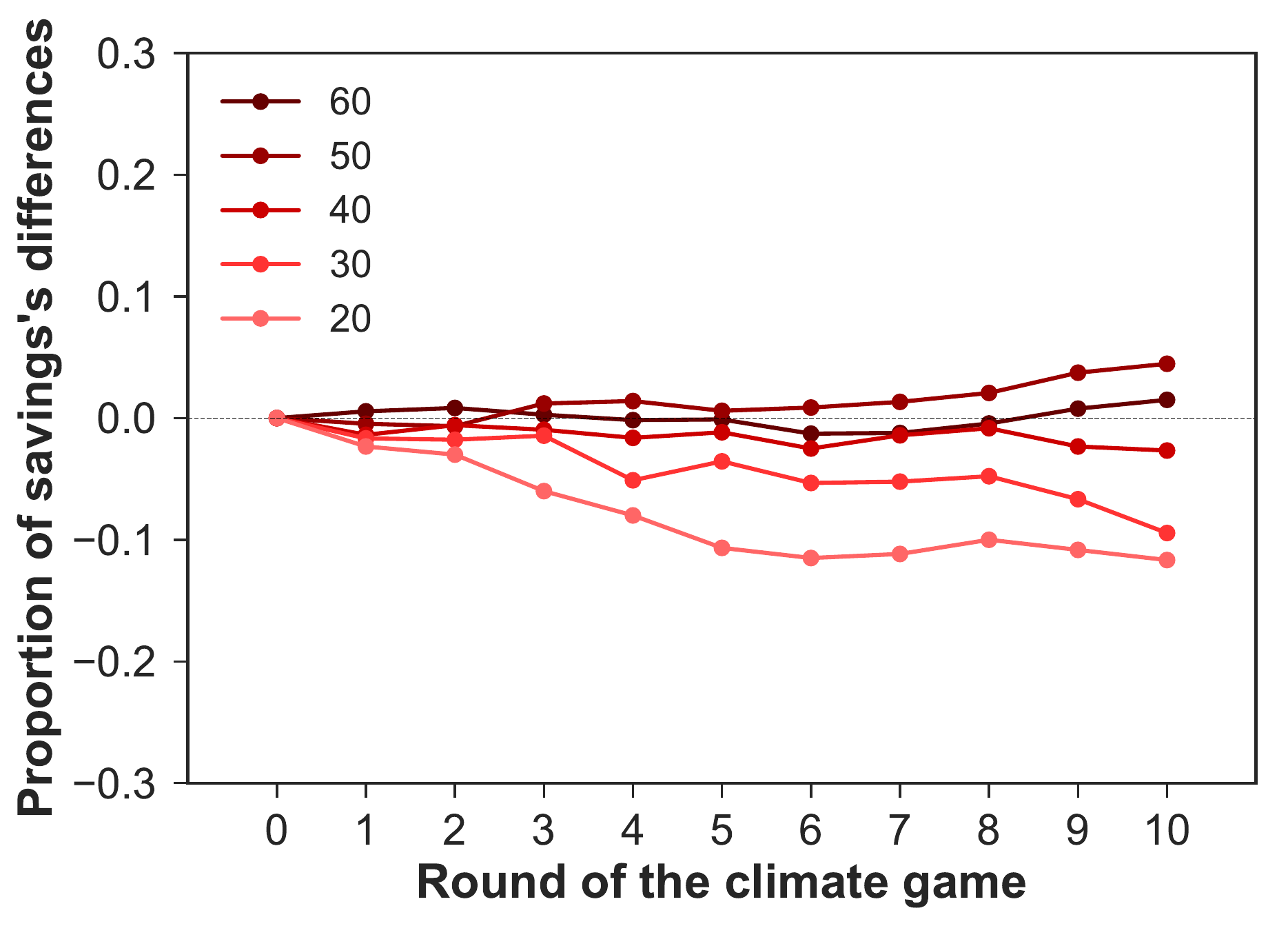}

\end{center}
\caption{(Left) Proportion of savings (mean and sem) at the end of the game per endowment and investment treatment. (Right) Differences of remaining capital $-$savings (S)$-$ between the treatment 01234 and 024 per endowment ($S_{01234}-S_{024}$ per endowment in each round).}
\label{fig:capital_evolution}
\end{figure}

The proportion of capital saved in both treatments have no significant difference (t-test p < 0.05) if we compare participants with same endowments (see Fig. \ref{fig:capital_evolution}(left)). However we observe in the evolution of the average capital grouped by endowments how adding more options creates a significantly larger gap between the average contribution of richer (50-60 \euro{}) and poorer participants (20-30 \euro{}). It is also interesting to observe that while the gap is created in the first five rounds, it keeps getting larger until the end of the game (see Fig. \ref{fig:capital_evolution}(right)).


\section*{Game evolution}

\begin{figure}[ht]
\begin{center}
\includegraphics[width=0.4\textwidth]{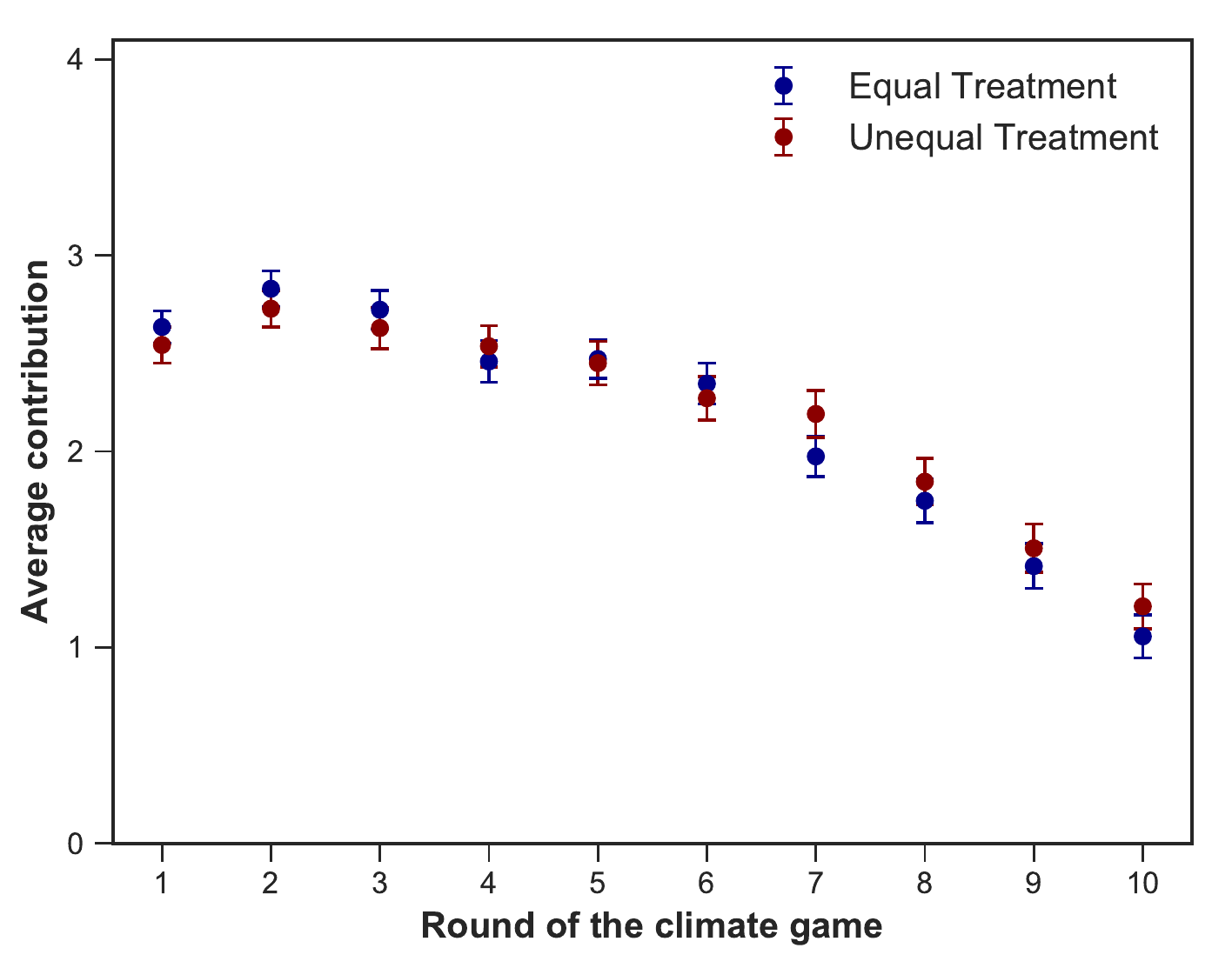}
\caption{Average individual investment and standard error of the mean by treatment over the game evolution. In both Equal Treatment and Unequal Treatment, participants's contribution decreases along the game. The differences between the two treatments are not statistically significants (MWU Two-Sided, U: 50.0, P: 0.97).}
\label{fig:game_evolution_round}
\end{center}
\end{figure}

In our experiment we collected how much one subject contributed in each round (Fig. \ref{fig:game_evolution_round}). In order to study and infer the existence of behavioral groups or patterns in the data, we first normalized the data so as to find the best classification.


First of all, we normalized the contribution done in each round based on the initial endowments. This is needed in order to compare meaningfully contributions of subjects in the unequal treatment with low endowments (20-30 \euro{}), mid endowments (40\euro{}), and high endowments (50-60 \euro{}), and all of them with the subjects of the equal treatment where all subjects have the same endowment (40\euro{}). Note that with this procedure, a normalized contribution of 0.5 represents the fair contribution in all treatments and all endowments.

By looking at the evolution of the game, we observe that each group reaches the target in different rounds $-$between the 6th and the 10th round (see Fig. \ref{fig:distribution_achieved_goals}). Therefore, to make the contributions comparable we decided to analyze the evolution of the game not according to the contributions per round, but instead by calculating the average contribution in regard to the accumulated capital in the pot at the beginning of the round. 


\subsection*{Phases of the game} 
\begin{figure}[ht]
\begin{center}
\includegraphics[width=0.4\textwidth]{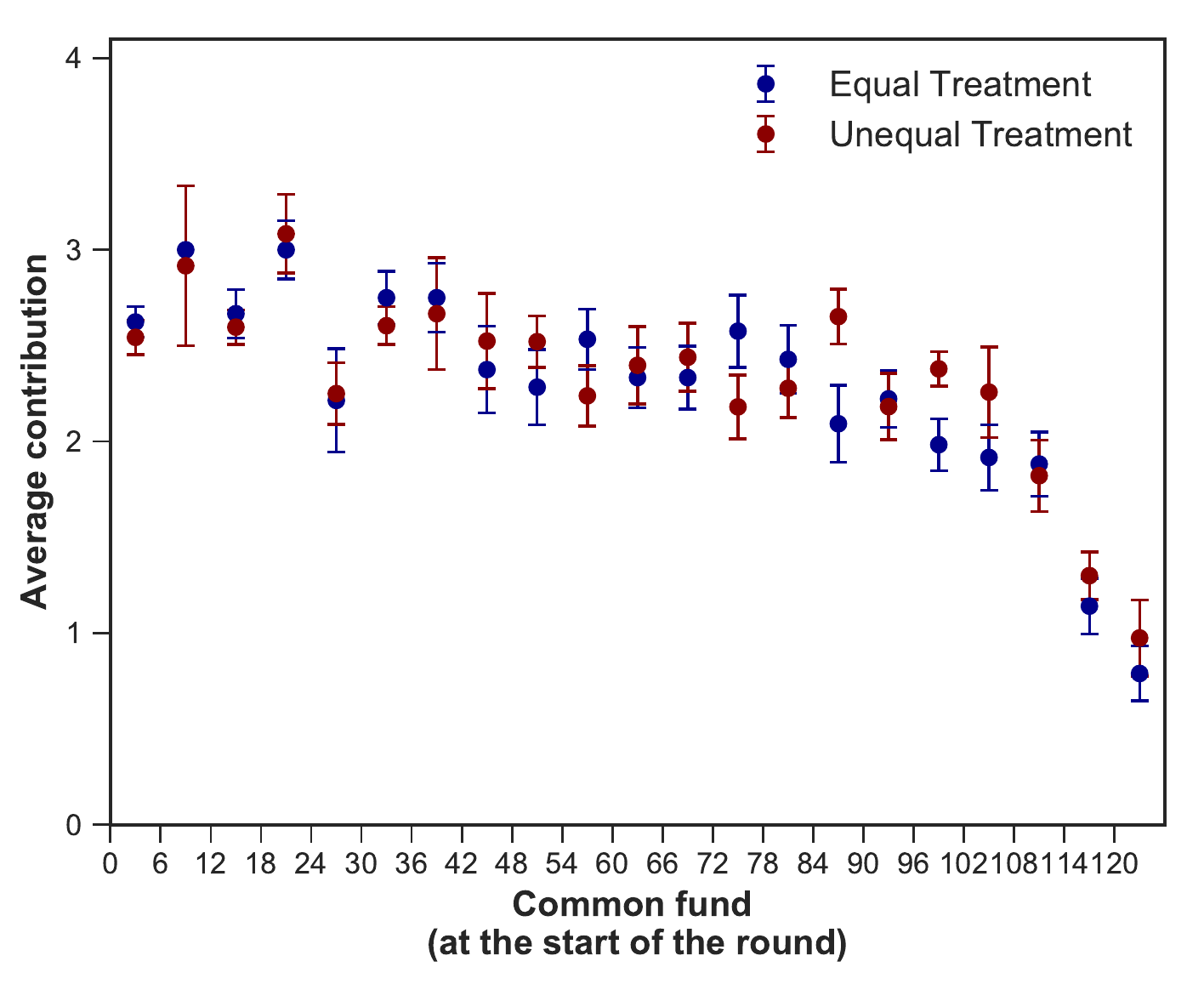}

\caption{Average individual investment and standard error of the mean by treatment over the game evolution. Decisions are grouped according to the total capital invested on the common fund at the start of the round. In both Equal Treatment and Unequal Treatment participants contribute above the fair contribution in the first part of the game and decrease when they are close to reach the target. We can observe three different regions on the game evolution: first, from 0-30 \euro{} participants are more erratic and at the same time contribute more to the average value. Second, from 30 to 90 \euro{} approximately there is a stable contribution slightly above the ideal average contribution. And third, after 90 \euro{} and until the goal is reached participants decrease substantially their final contribution.}
\label{fig:game_evolution_bins}
\end{center}
\end{figure}

\begin{figure}[ht]
\begin{center}
\includegraphics[width=0.3\textwidth]{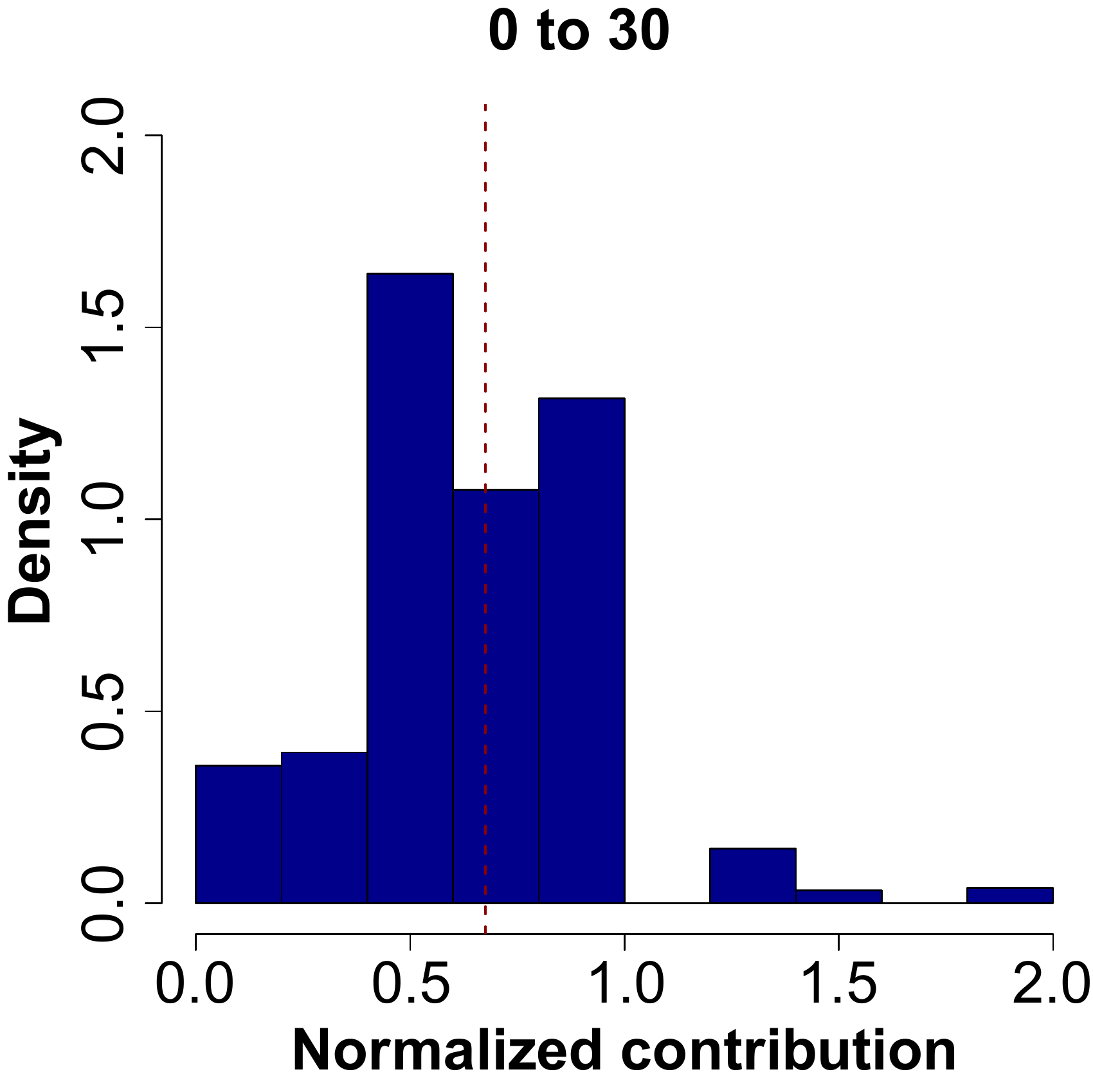}
\includegraphics[width=0.3\textwidth]{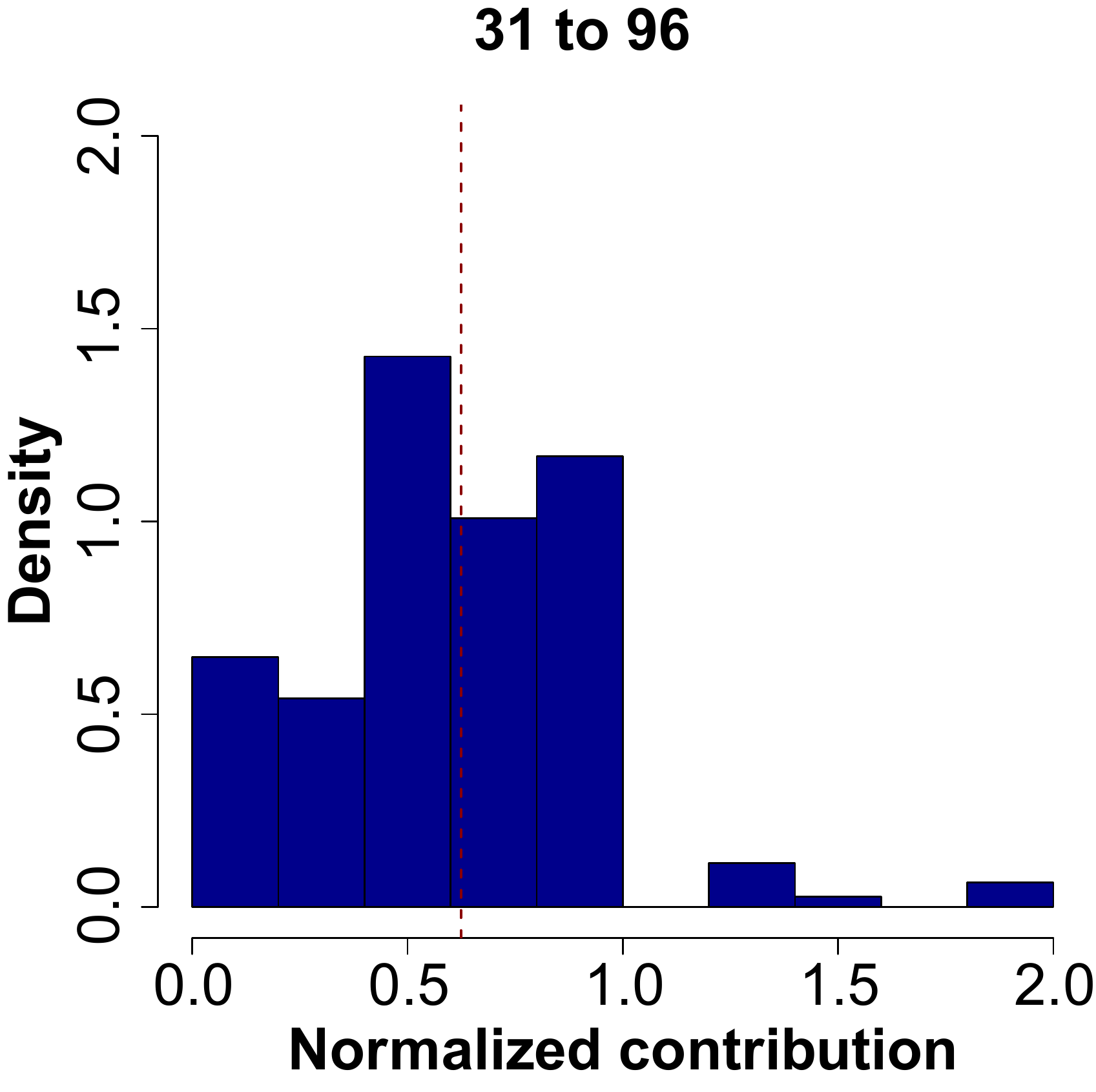}
\includegraphics[width=0.3\textwidth]{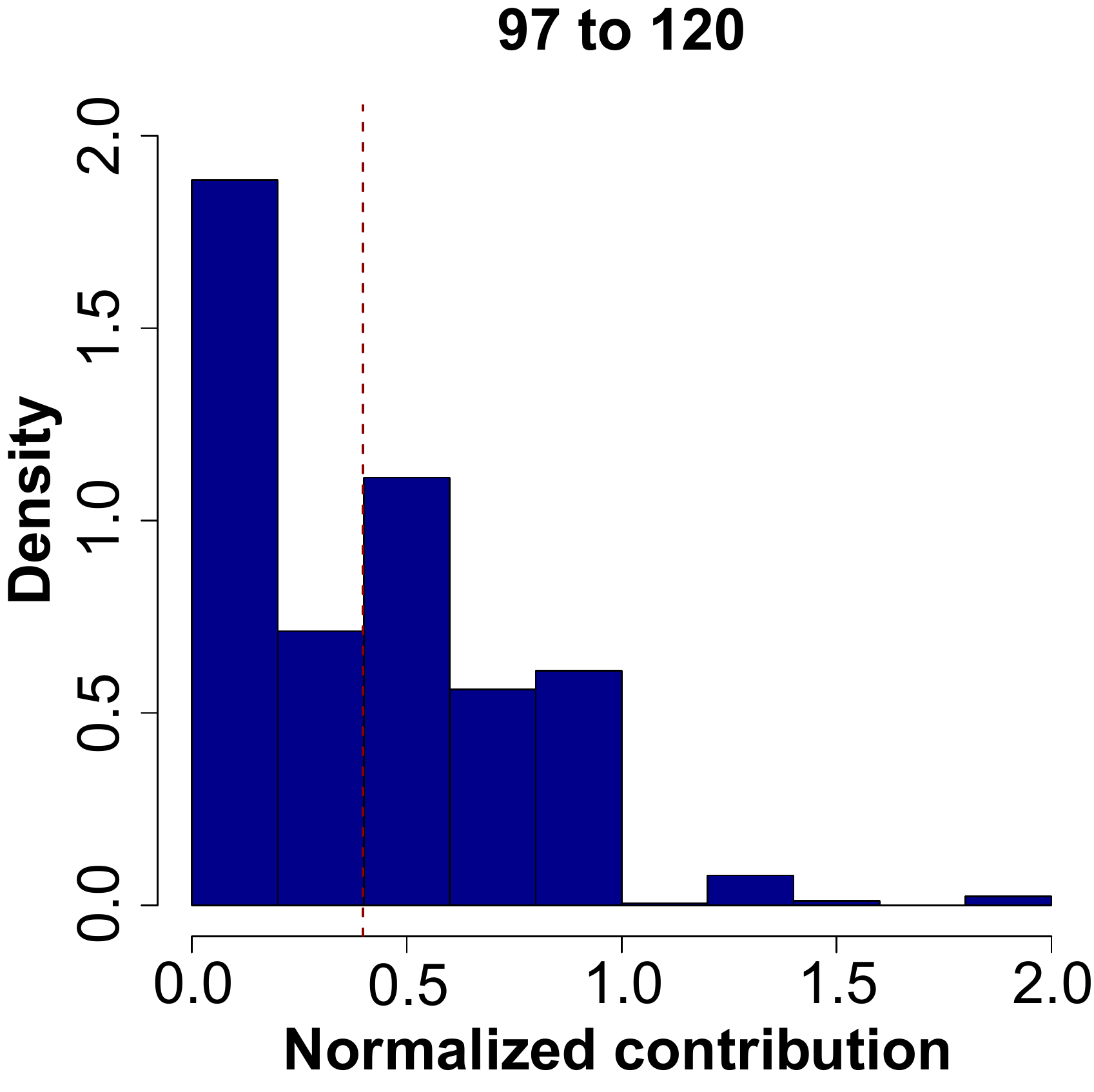}
\caption{Distributions of normalized contributions in the three phases of the game. The mean (SD) in each phase, based on the accumulated capital in the common fund, is: common fund from 0 to 30 \euro{}: 0.67 (0.33), common fund from 31 to 96 \euro{}: 0.62 (0.37), and common fund from 97 to 120 \euro{}: 0.39 (0.38).}
\label{fig:distribution_contributions_phases}
\end{center}
\end{figure}

The analysis of the evolution of the game according to the amount contributed to the common pot provides a better description of the behavior of the groups, and allows distinguishing different phases of the game. For instance (see Fig. \ref{fig:game_evolution_bins}), we have observed that at the beginning of the game the subjects contribute a lot and explore the contributions of the other participants. However, when they get closer to the goal, there is an important change on how people contribute: their contributions decrease when they see that the goal is close to being achieved (see Fig. \ref{fig:distribution_contributions_phases}). 

\subsection*{Binning} To study the phases of the game, that is, the evolution independently of the rounds, we bin the rounds according to the accumulated capital in the common fund, and calculate the participants's average contribution per bin and normalize it (see Table \ref{tab:binning} for an example on how this normalization is done).


\begin{table*}[tbhp!]
\centering
\caption{Example of user's contribution normalization and binning in a particular game.}
\label{tab:binning}
\begin{tabular}{lcccccccccc}
Round & 1 & 2 & 3 & 4 & 5 & 6 & 7 & 8 & 9 & 10\\
\midrule
Contribution$^{\text{1}}$ & 4 & 3 & 4 & 3 & 2 & 2 & 4 & 3 & 3 & 0\\
Remaining to target$^{\text{2}}$ & 120 & 102 & 82 & 61 & 49 & 26 & 21 & 8 & 1 & -5\\
Common Fund$^{\text{3}}$ & 0 & 18 & 38 & 59 & 71 & 84 & 99 & 112 & 119 & 125\\
\bottomrule
Binning (bin=24)$^{\text{4}}$ & \multicolumn{2}{c}{0-23} & 24-47 & \multicolumn{2}{c}{48-71} & 72-95 & \multicolumn{3}{c}{96-119} & $\geqslant120$ \\
\midrule
Av. Cont.$^{\text{5}}$ & \multicolumn{2}{c}{3.5} & 4 & \multicolumn{2}{c}{2.5} & 2 & \multicolumn{3}{c}{3.3} & 0 \\
Av. Cont. Norm.$^{\text{6}}$ & \multicolumn{2}{c}{0.875} & 1 & \multicolumn{2}{c}{0.625} & 0.5 & \multicolumn{3}{c}{0.83} & 0 \\
\bottomrule
\end{tabular}

\vspace{1ex}
	\raggedright $^{\text{1}}$ Contribution of a single user over the game (10-rounds). $^{\text{2}}$ Capital remaining to achieve the goal in a particular game (120\euro{} at the beginning of the game). $^{\text{3}}$ Capital contributed and accumulated in each round of a particular game. $^{\text{4}}$ Binning the common fund in groups of 24. $^{\text{5}}$ Average contribution of a single user in the bin. $^{\text{6}}$ Average contribution normalized of a single user in the bin..

\end{table*}



\section*{Individual Behavior}
\begin{figure}[!h]  
\begin{center}
\includegraphics[width=0.4\textwidth]{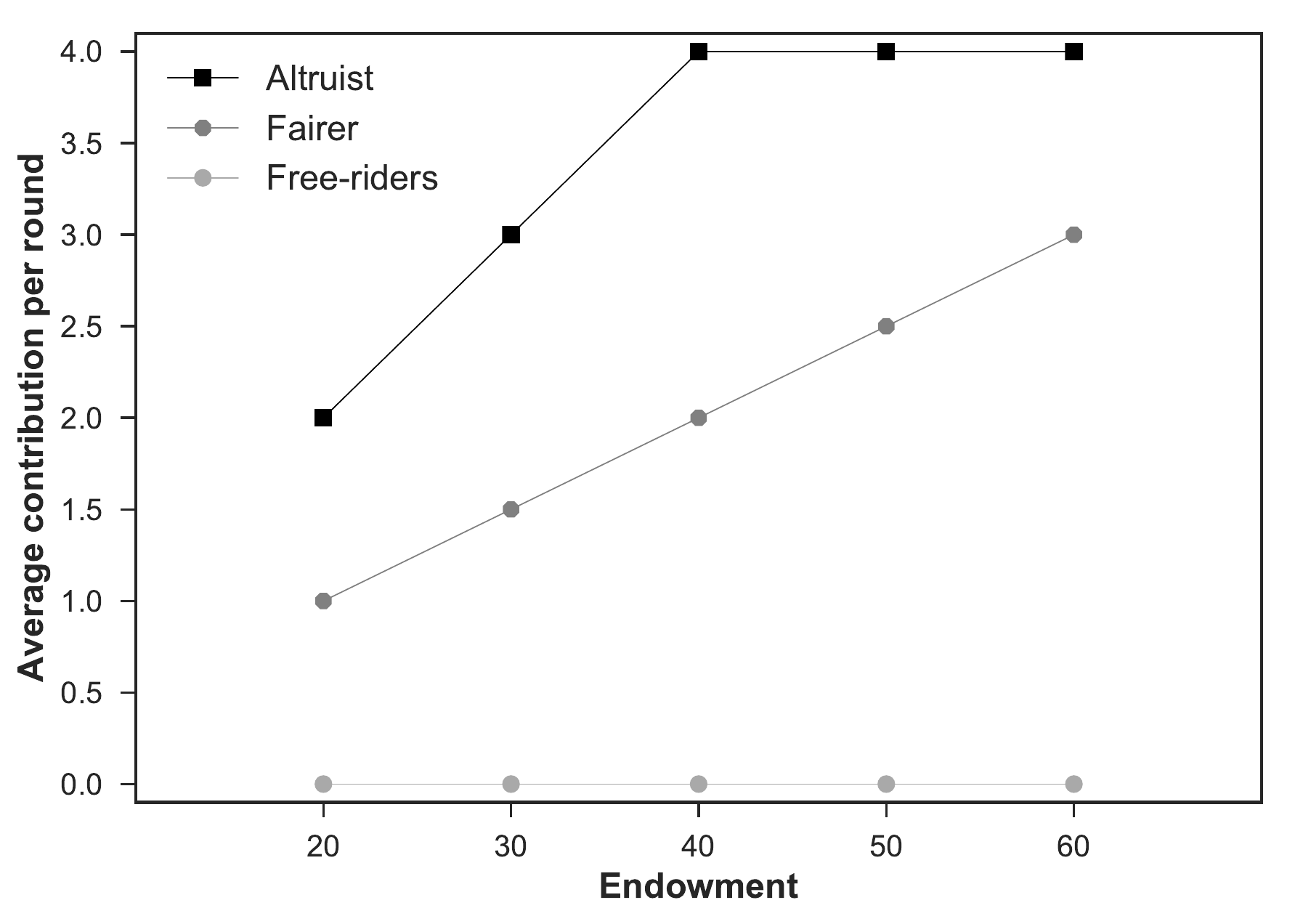}
\end{center}
\caption{Ideal "pure" strategies based on our experiment design.}
\label{fig:pure_strategies}
\end{figure}

To study the contribution strategies of the participants of our experiment we first checked whether they follow "pure" strategies by looking at the total of their contributions over the whole game. Note that we only take into account the contributions before the target was reached, and also that the strategies are conditioned by the inequality of endowments. We consider three "pure" strategies: (i) free-riders, those with a contribution of 0\euro{}; (ii) fairers, those that have an average contribution per round of 1\euro{} (20\euro{}), 1.5\euro{} (30\euro{}), 2\euro{} (40\euro{}) and so on; and (iii) altruists, those who contribute the maximum that it is possible according to their initial capital, with average round contributions of 2\euro{}, 3\euro{}, 4\euro{}, 4\euro{} and 4\euro{} per endowments of 20\euro{}, 30\euro{}, 40\euro{}, 50\euro{} and 60\euro{} respectively (see Fig. \ref{fig:pure_strategies}). 
We observe that the subjects rarely follow those 'pure' strategies, only 42 out of 320 subjects followed a particular strategy: 1 free-rider, 30 fairers and 11 altruists.

In light of the above results and the difficulties to find clear strategies in the experiment, we next ran an unsupervised algorithm to understand the variety of subject's behavior over the game.

\subsection*{Clustering Analysis}
\begin{figure}[!h]  
\begin{center}
\includegraphics[width=0.4\textwidth]{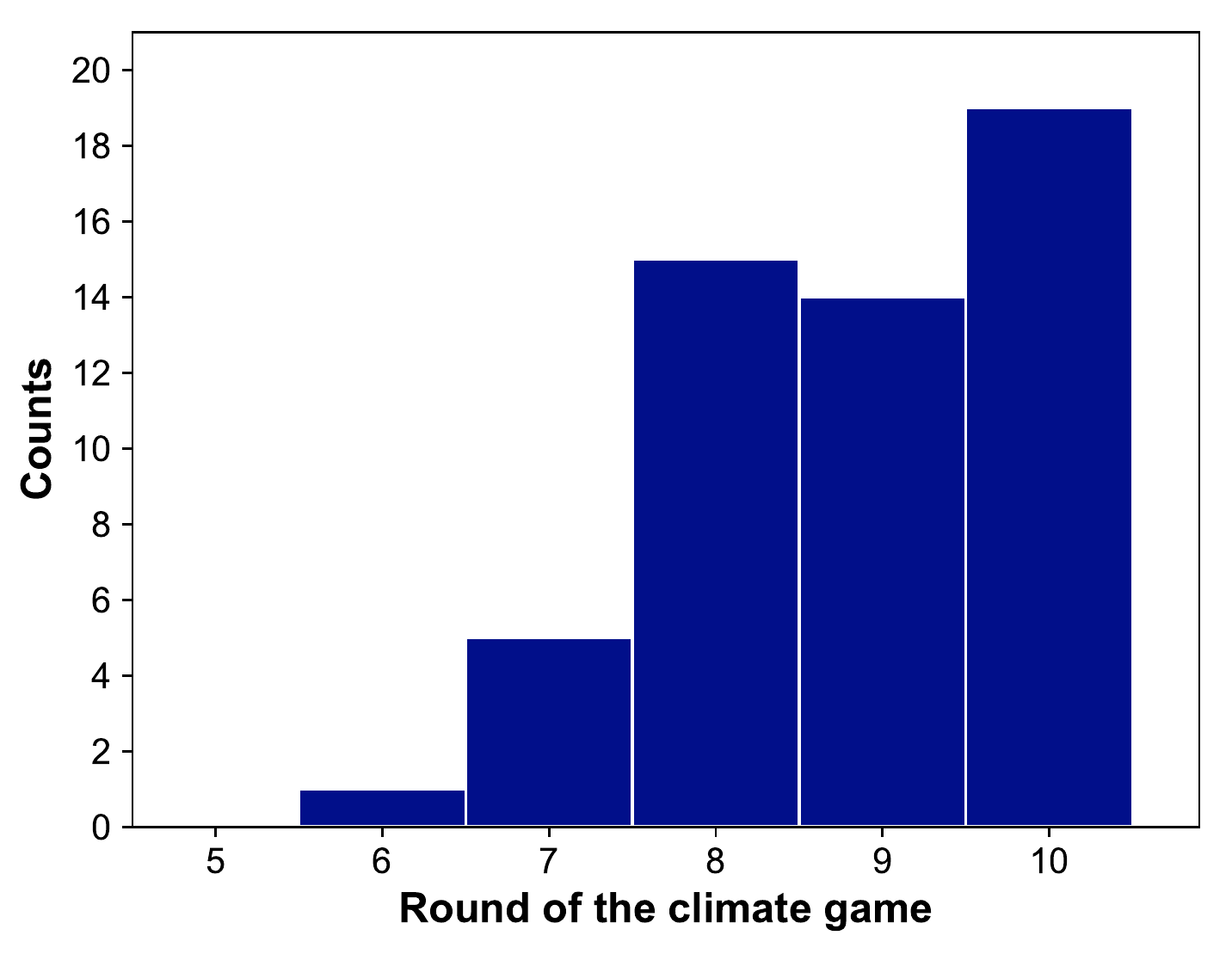}
\end{center}
\caption{Number of games in which the goal has been achieved in a particular round. The average (SD) round is 8.83 (1.07).}
\label{fig:distribution_achieved_goals}
\end{figure}

We hypothesized that there exist different strategies of cooperation in our dataset that can not be described in terms of pure strategies due to their complexity, but that they could be revealed using unsupervised learning techniques. Hence, we ran hierarchical cluster algorithm on our data to analyze the structure of subject's contributions. We represent the sequence of contribution as explained above (see Table \ref{tab:binning}), creating a matrix of contribution, where every row represents a vector of individual contributions over the game. We use the accumulated capital in the common fund instead of rounds because every game could finish in a different round (see Fig. \ref{fig:distribution_achieved_goals}), and a bin = 24 due to the rapid evolution of the game. Therefore, in each cell, we have the average contribution normalized (regarding the endowment) in the round that grouped every bin.

We implemented an agglomerative hierarchical clustering strategy (using the hclust package in R) to find an initial approximation of groups using ward.D2 and euclidean distances. The agglomerative strategies consist of a "bottom-up" approach, each user contribution starts in its own cluster and pairs of clusters are merged in each step based on the optimal value of an error sum of squares (Ward's method). With the objective to determine the number of groups that better fits with our data, we ran an algorithm (NBCluster package in R) that computes 26 indexes and recommends the optimal number of clusters according to the majority rule.


Once we know that there is sufficient evidence to find groups of participants with different strategies and behavior, and to ensure that the clustering results are robust and reliable we ran an implementation of consensus clustering (ConsensusClusteringPlus package in R). Consensus clustering 
determines the number of clusters and computes consensus values such as item consensus and cluster consensus. Item consensus represents the membership of an item with all items in a particular cluster, this value indicates if a particular item is a pure member of the cluster or if it is unstable. Cluster consensus provides information about the consensus between members of a group, high values indicate high stability.

The parameters we used to perform the calculation illustrated in the next sections are: maximum evaluated k of 9 so that groups count of 2 to 9 are evaluated; 1000 re-samplings, agglomerative hierarchical clustering algorithm, euclidean distances and ward.D2 linkage. 

\subsection*{Equal Treatment}
The equal treatment of the experiment includes 162 subjects (27 games), of which 159 subjects contributed with valid actions. In order to analyze their individual strategies we created a matrix with the average contribution in each stage of the game binned by the accumulated capital in the common fund, 
and then we computed the cluster consensus in our equal treatment dataset.

\begin{figure}[ht]
\begin{center}
\includegraphics[width=0.65\textwidth]{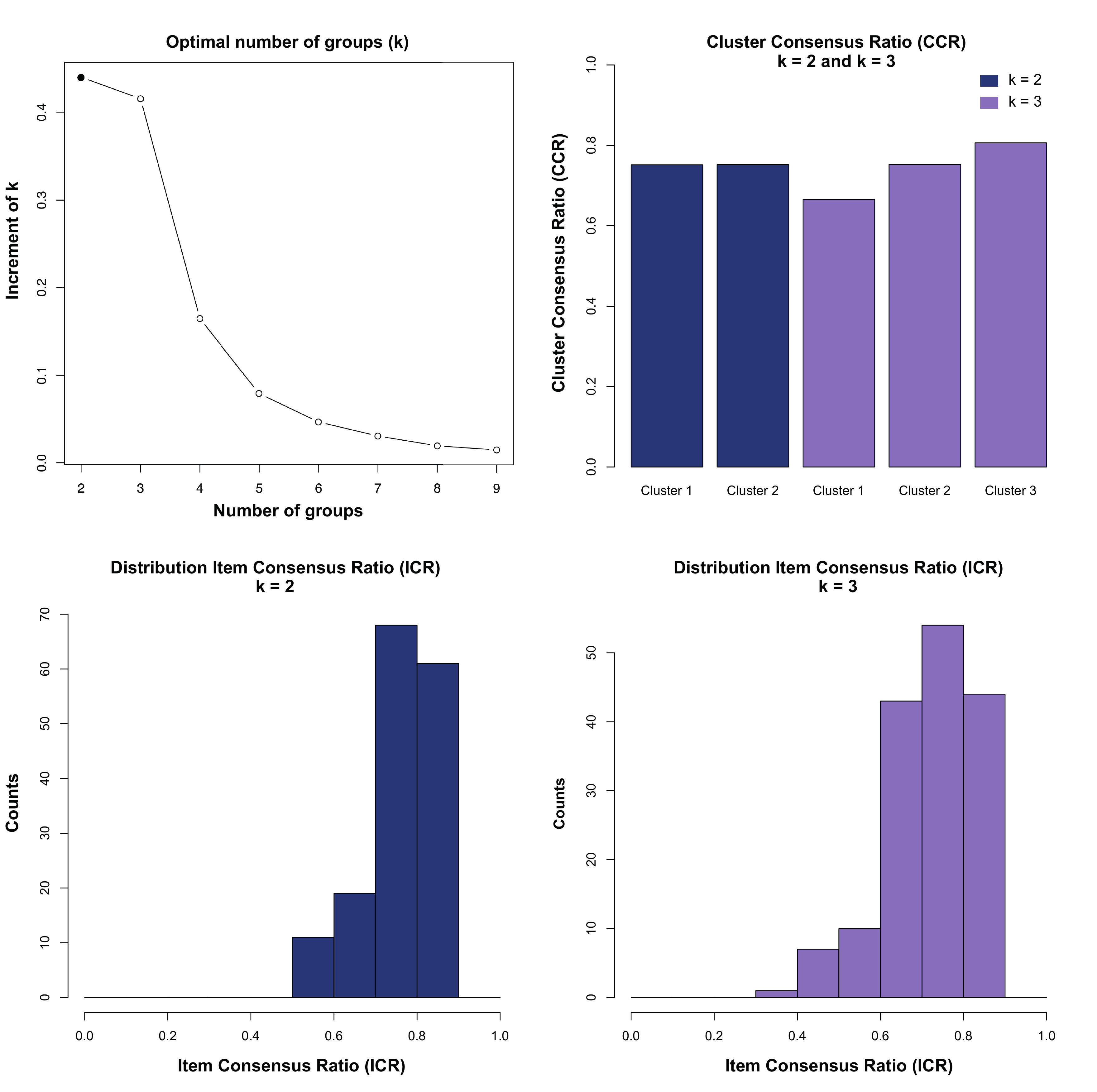}
\caption{Equal Treatment. (Top-Left) Optimal number of clusters. (Top-Right) Cluster consensus ratio. (Bottom) Item consensus ratio.}
\label{fig:panel_equal}
\end{center}
\end{figure}
\begin{figure}[ht]
\begin{center}
\includegraphics[width=0.3\textwidth]{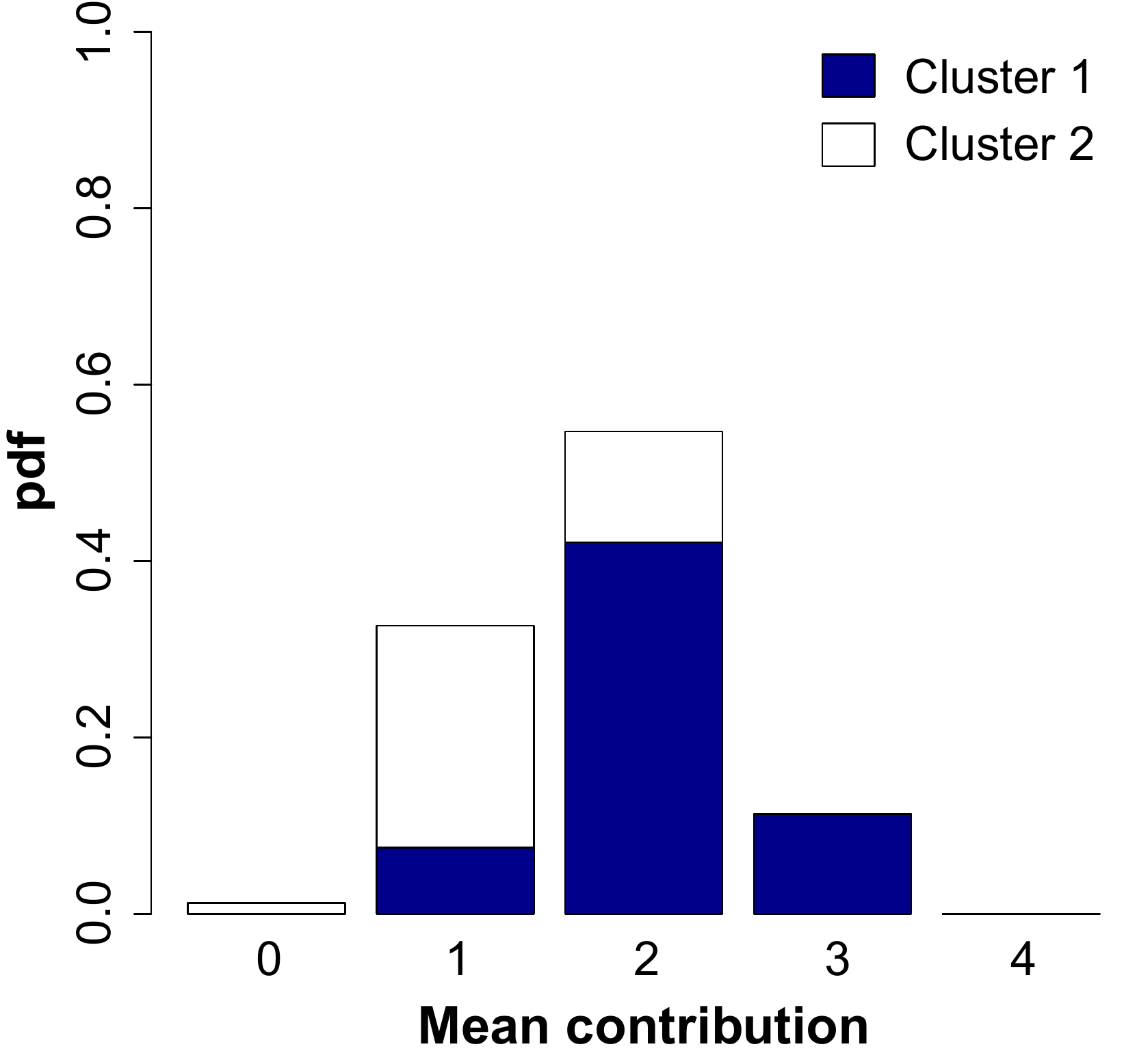}
\includegraphics[width=0.3\textwidth]{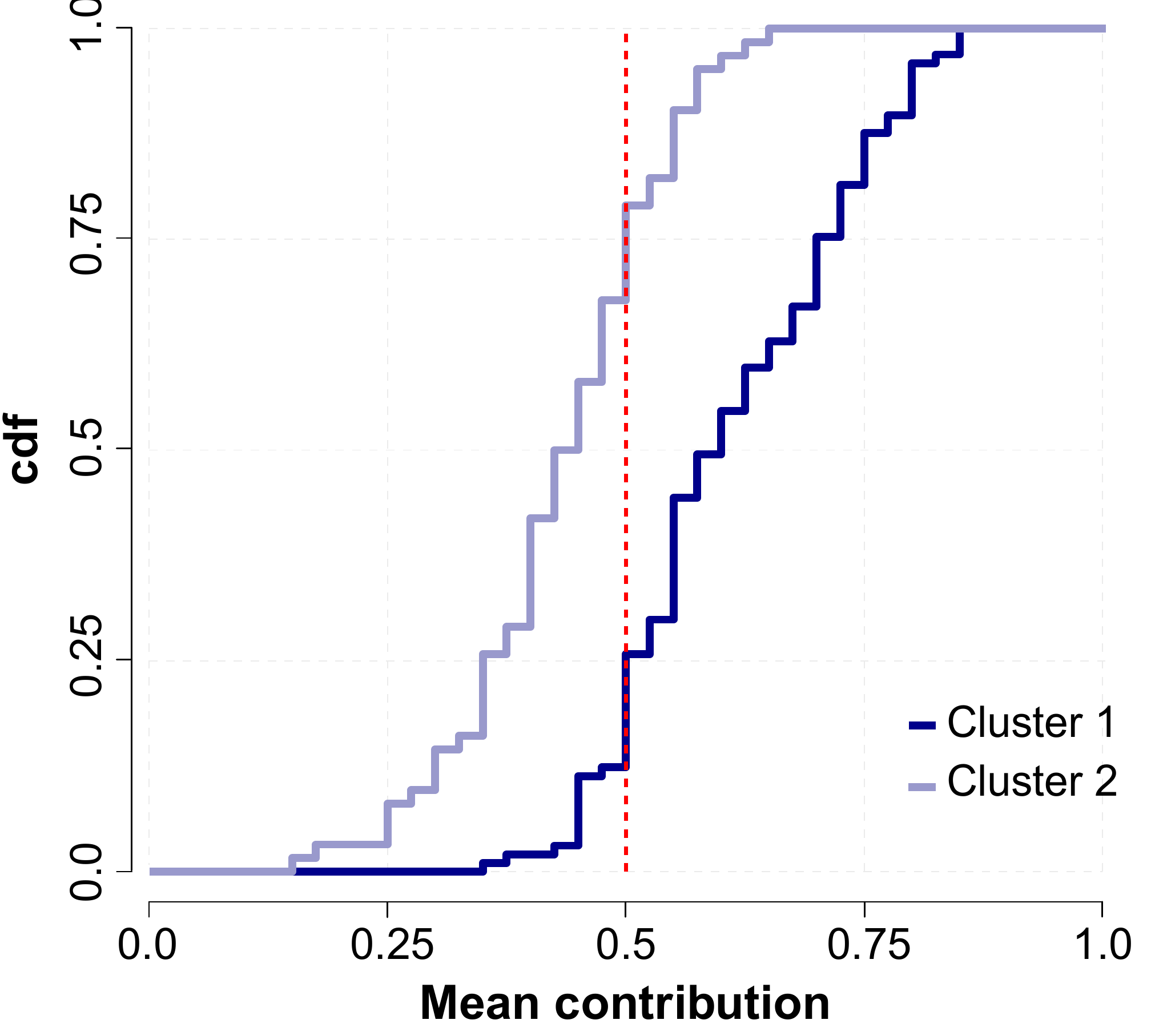}
\caption{(Left) Distribution of subjects in clusters based on their average contribution per round. (Right) Cumulative distribution function based on their average contribution per round.}
\label{fig:distributions_equal}
\end{center}
\end{figure}

The number of clusters that better fits with our data and maximizes the consensus values is 2 (see Fig. \ref{fig:panel_equal}). The average (SD) consensus clustering ration for 2 groups is 0.75 (0) and the average of item consensus is 0.75 (0.08). Cluster 1 is composed by 97 subjects (61\%) while cluster 2 is formed by 62 subjects (39\%). Fig. \ref{fig:distributions_equal} represents the distribution of subjects (pdf) and the cumulative distribution function (cdf) of their average contribution in both clusters.


\subsection*{Unequal Treatment}

The unequal treatment includes 162 subjects (27 games), of which 161 contributed valid data. The matrix of contributions was formed in the same way that in the equal case.

\begin{figure}[ht]
\begin{center}
\includegraphics[width=0.5\textwidth]{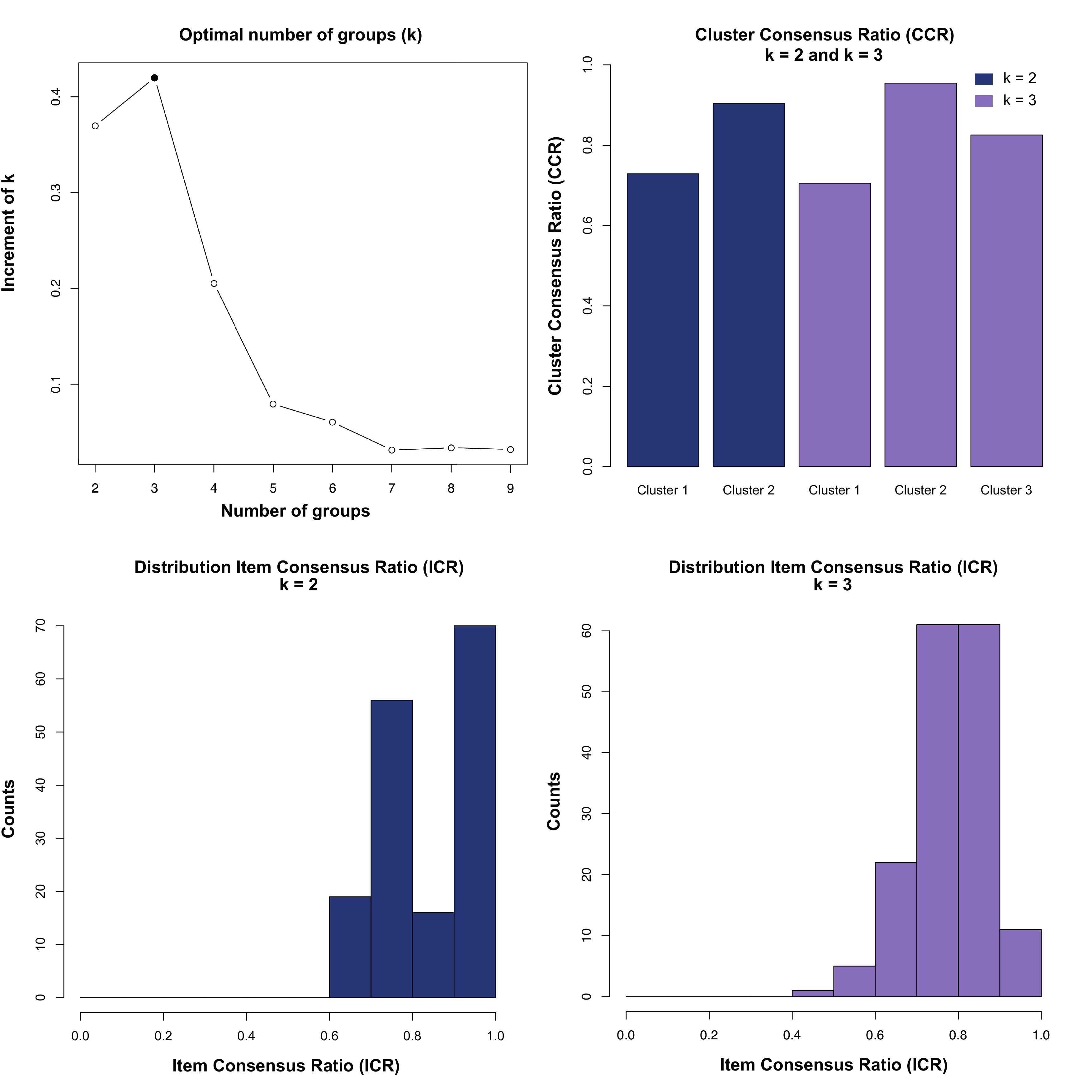}
\caption{Unequal Treatment. (Top-Left) Optimal number of clusters. (Top-Right) Cluster consensus ratio. (Bottom) Item consensus ratio.}
\label{fig:panel_unequal}
\end{center}
\end{figure}

The consensus cluster approach concluded that 3 clusters is the most stable number of groups in this treatment, see Fig. \ref{fig:panel_unequal}. In this case, the average (SD) consensus clustering ratio is 0.83 (0.12) and the average of item consensus ratio is 0.78 (0.09).

\begin{figure*}[ht]
\begin{center}
\includegraphics[width=0.99\textwidth]{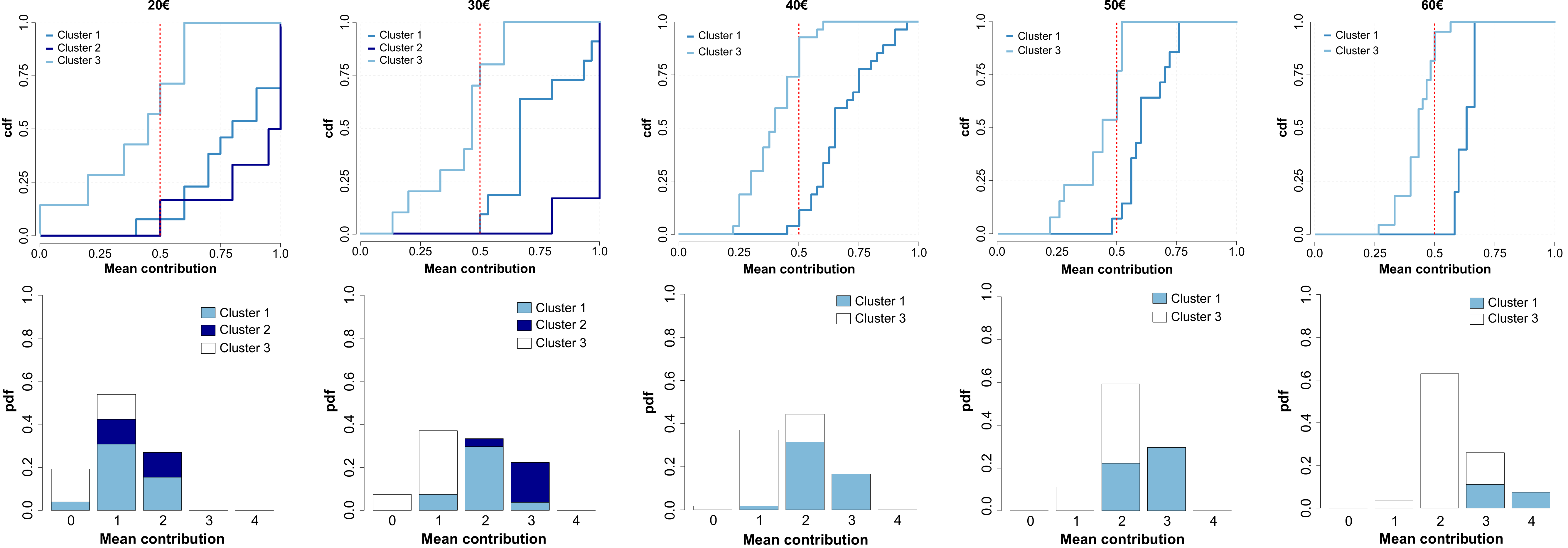}
\caption{(Top) Cumulative distribution function based on their average contribution per round. (Bottom) Distribution of subjects in clusters based on their average contribution per round.}
\label{fig:distributions_unequal}
\end{center}
\end{figure*}

In contrast to the equal treatment, in the unequal scenario a new group appears. The composition of clusters is: cluster 1, 70 subjects (43.48\%); cluster 2, 12 subjects (7.45\%); and cluster 3, 79 subjects (49.07\%). The most populated cluster is the one composed by subjects that contributed around and below the fair contribution. Fig. \ref{fig:distributions_unequal} represents the distribution of subjects (pdf) and the cumulative distribution function (cdf) of their average contribution in both clusters.


\section*{Capital Inequality}
\begin{figure}[ht]
\begin{center}
\includegraphics[width=0.7\textwidth]{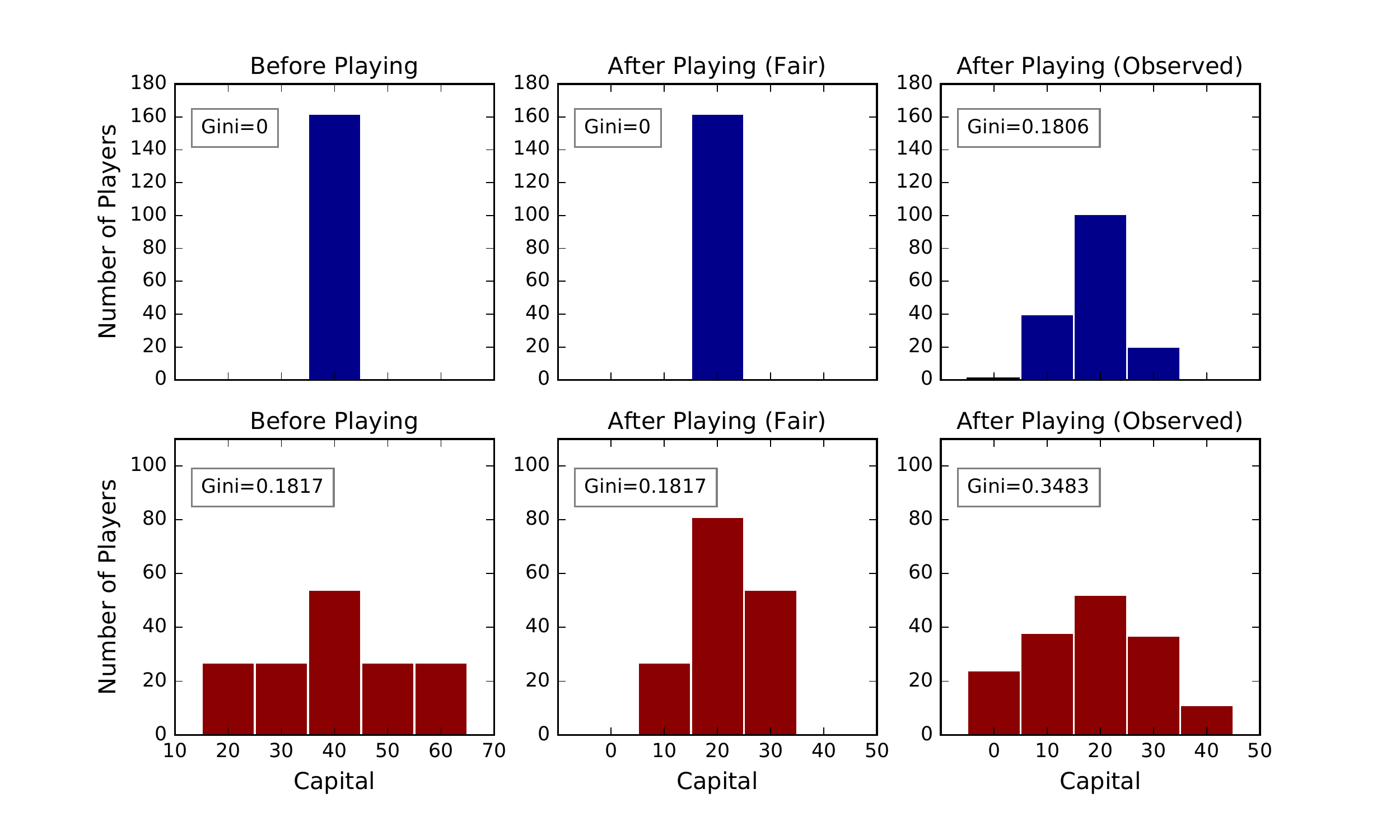}
\caption{Distribution of capitals in different scenarios of the game for the equal treatment (first row in light blue) and the unequal treatment (bottom row in red). First column displays the corresponding distribution of endowments at the beginning of the game. Second column pictures the hypothetical final distribution if all the participants contributed fairly, i.e. with half of their endowments. Finally in the third column, the observed final distribution of capitals is shown. The Gini coefficient is indicated in every case.}
\label{fig:gini}
\end{center}
\end{figure}

As shown throughout the main text, participants with less initial capital tend to contribute much more in relative terms. As a result of this behavior, we also observe an increase of inequality that we would not observe if the participants contributed the amount defined as fair in the main text (see Fig. \ref{fig:gini}). We also measure such inequality by means of the Gini Coefficient, a standard measure in economics used extensively for studying income distributions and its inequality.

\section*{Earnings}

\begin{figure}[ht]
\begin{center}
\includegraphics[width=0.5\textwidth]{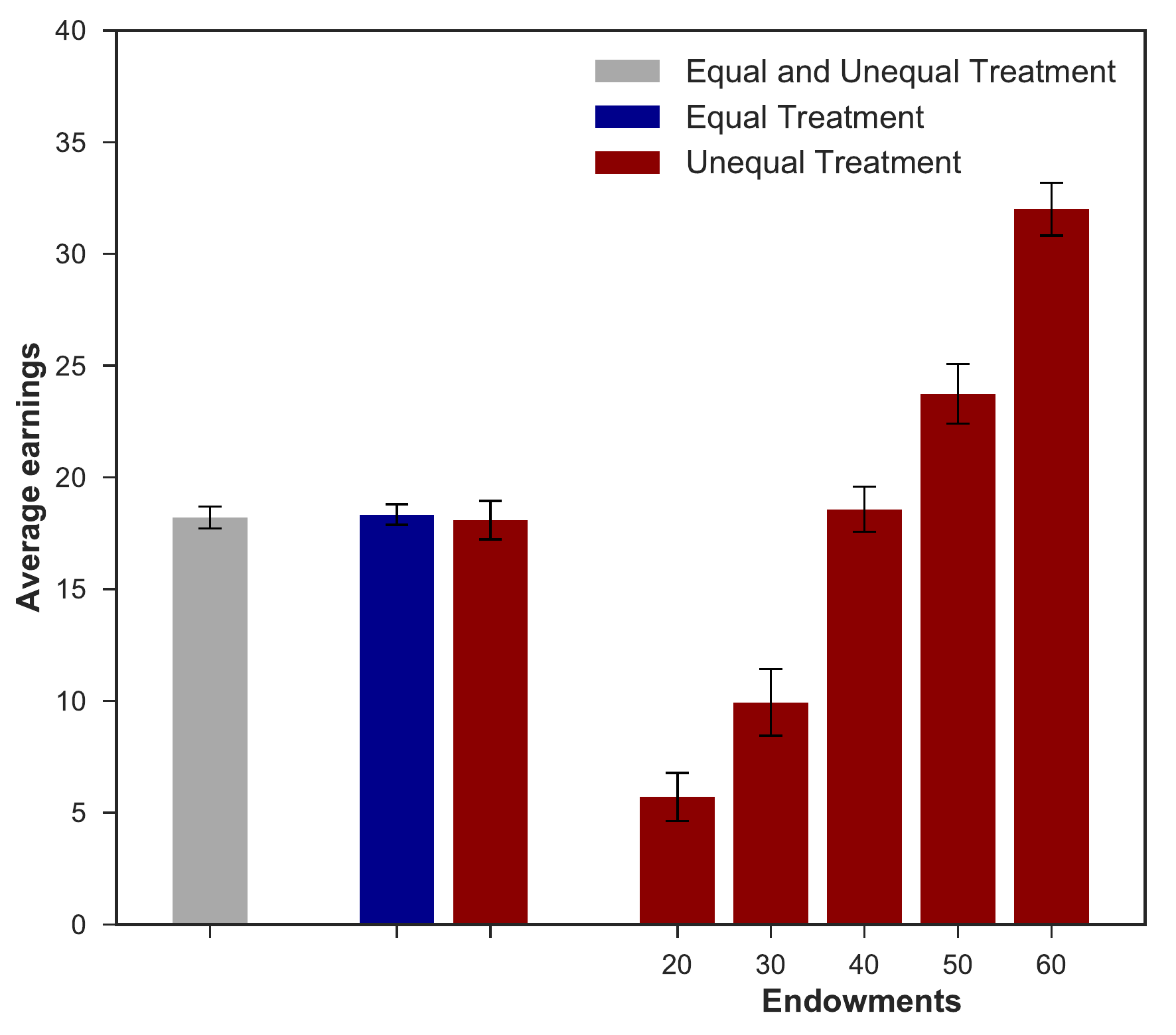}
\caption{Average earnings and standard error of the mean regarding treatment and endowments}
\label{fig:earnings}
\end{center}
\end{figure}

Once the game finished and if the subjects achieved the goal, they kept the capital not contributed and received it in the form of a gift card. The average (standard deviation) earnings among all the subjects were 18.21\euro{} (8.8). There exist no significant differences in the means (t-test; t:0.26, p:0.8) between the equal treatment 18.33\euro{} (5.8) and the unequal treatment 18.08\euro{} (11). Fig. \ref{fig:earnings} illustrates earnings regarding of endowments and treatments in detail.

\section*{Decision Making Times}
\begin{figure}[ht]
\begin{center}
\includegraphics[width=0.4\textwidth]{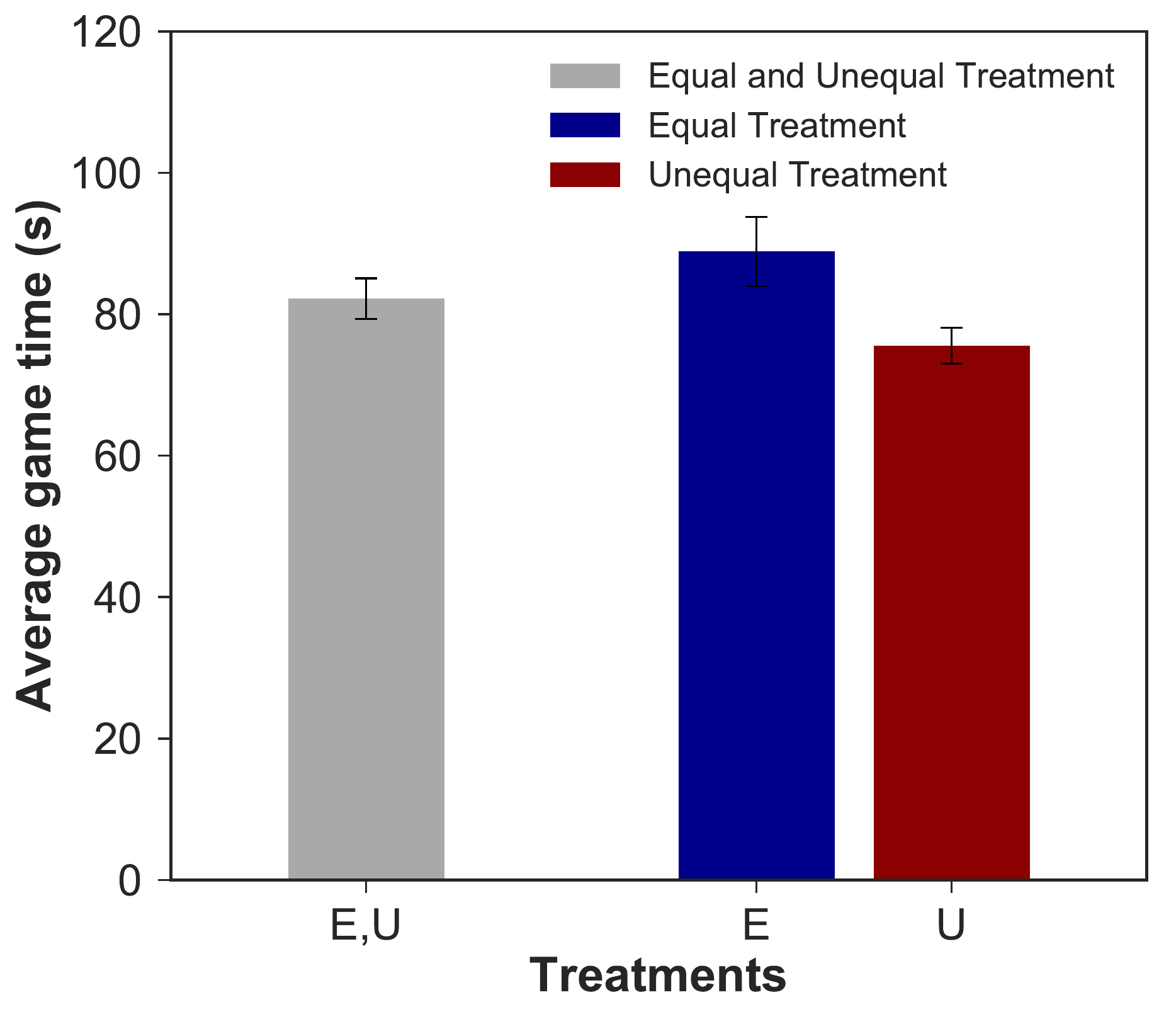}
\includegraphics[width=0.4\textwidth]{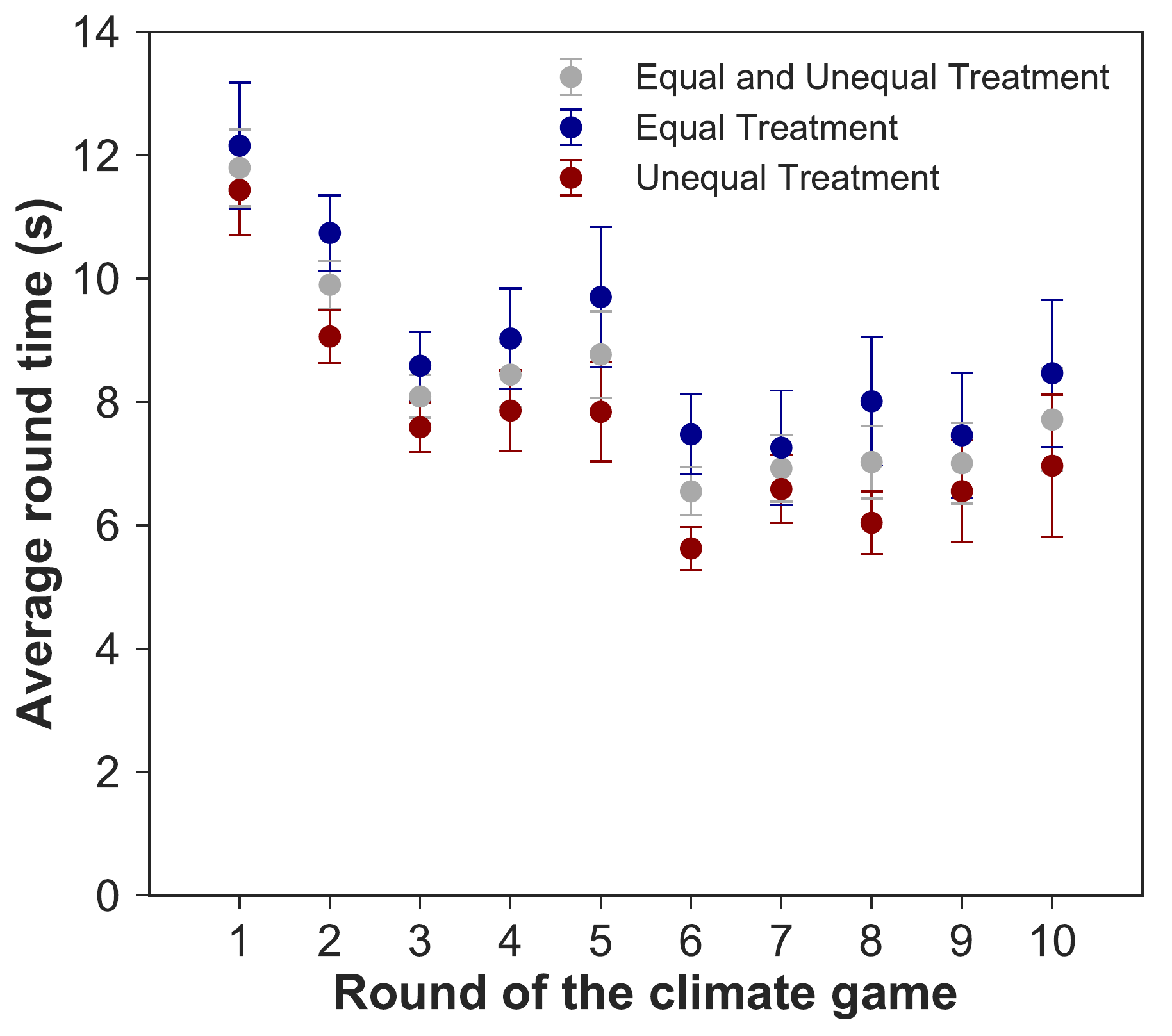}
\caption{(Left) Duration of a game (mean and sem) per treatment. (Right) Evolution of decision making times over round.}
\label{fig:decision_making_times}
\end{center}
\end{figure}

The platform designed to carry out the experiment also records the decision making times and the duration of a round, and we can also obtain the total playing time by suming the durations of rounds in a game. 
The average (standard deviation) duration of all games was 82.21s (20.98), and the average (standard deviation) duration of a round was 8.22s (4.47). There are significant differences in the mean (t-test; t:2.41, p:0.019) between equal treatment 88.88s (25.01) and unequal treatment 75.55s (12.89). The decision times decrease as rounds go on, especially in the first five rounds, and subsequently stabilizes (see Fig. \ref{fig:decision_making_times}).

\section*{Questionnaire}
After the collective-risk dilemma ended, as a last stage of the experiment, for each group of subjects we asked a questionnaire with two sets of questions: (i) how they made decisions in the game and (ii) some basic concepts about climate change. The questionnaire was also presented identically in three languages (Catalan, Spanish and English), here we present  the questions in English.

Question 1: \textit{Had you previously participated in Citizen Science experiments?} Answers: a.\textit{No}; b.\textit{Yes}; c.\textit{Yes, in previous DAU experiments}.

Question 2: \textit{Did you like the experience?} Answers: a.\textit{Very much}; b.\textit{Somewhat}; c.\textit{Not really}; d.\textit{Not at all}.

Question 3: \textit{At the beginning of the experiment, did you expect to reach the 120\euro{} target?} Answers: a.\textit{Yes, from the beginning}; b.\textit{Yes, after a few rounds}; c.\textit{No, from the beginning}; d.\textit{No, after a few rounds}.

Question 4: \textit{Generally, if others are contributing little, I should also contribute little.} Answers: a.\textit{Agree}; b.\textit{Disagree}; c.\textit{My contribution should not depend on this}; d.\textit{n/a}.

Question 5: \textit{Generally, if others are contributing a lot, I should also contribute a lot.} Answers: a.\textit{Agree}; b.\textit{Disagree}; c.\textit{My contribution should not depend on this}; d.\textit{n/a}.

Question 6: \textit{I think there have been players who have taken advantage of the generosity of others to maintain their capital.} Answers: a.\textit{Agree}; b.\textit{Disagree}; c.\textit{n/a}.

Question 7: \textit{The contributions of each player should be proportional to their capital: those who have started with more should contribute more and those who have started with less should contribute less.} Answers: a.\textit{Agree}; b.\textit{Disagree}; c.\textit{n/a}.

Question 8: \textit{The contributions of each player should be fair, because the benefits of the common fund affect everyone equally, as well as the risk of losing everything.} Answers: a.\textit{Agree}; b.\textit{Disagree}; c.\textit{n/a}.

Question 9: \textit{It seems fair to me that a player with a lot of capital should get money at the end, if he has contributed at least half of what he had.} Answers: a.\textit{Agree}; b.\textit{Disagree}; c.\textit{n/a}.

Question 10: \textit{When the polluting gases prevent the rays of the Sun from coming out of the Earth, it is due to...} Answers: a.\textit{Doppler Effect}; b.\textit{Greenhouse Effect}; c.\textit{Faraday Effect}; d.\textit{Refrigerator Effect}.

Question 11: \textit{Which of the following countries is the most polluting in the world?} Answers: a.\textit{U.S}; b.\textit{Italy}; c.\textit{China}; d.\textit{Japan}.

Question 12: \textit{Which of the following elements is the least polluting?} Answers: a.\textit{Oil}; b.\textit{Carbon}; c.\textit{Solar energy}; d.\textit{Nuclear energy}.

Question 13: \textit{Which international treaty tries to regulate CO2 emissions to the atmosphere?} Answers: 
a.\textit{Declaration of Helsinki}; b.\textit{Kyoto Protocol}; c.\textit{Schengen Agreement}; d.\textit{Treaty of Versailles}.

Question 14: \textit{What is the total number of gaseous pollutants emitted by each individual?} Answers: a.\textit{Carbon footprint}; b.\textit{Eco-Impact}; c.\textit{Individual gas fee}; d.\textit{Reduced environmental cost}.

Question 15: \textit{According to the economist Nicolas Stern, if urgent measures are not implemented, what costs could climate change represent in 2050, as a percentage of world GDP?} Answers: a.\textit{2\%}; b.\textit{5\%}; c.\textit{15\%}; d.\textit{20\%}

\section*{Tutorial}
Before the collective-risk dilemma started, we showed to each group of subjects a tutorial that was included in the same platform used to participate in the game. The subjects were assisted by researchers who answered any questions that came up about the experiment. The tutorial was presented identically in three languages (Catalan, Spanish and English). Here we show the tutorial screens and text of the English version.

\begin{figure*}[!th]  
\begin{center}$
\begin{array}{lll}
(a) \includegraphics[width=0.3\textwidth]{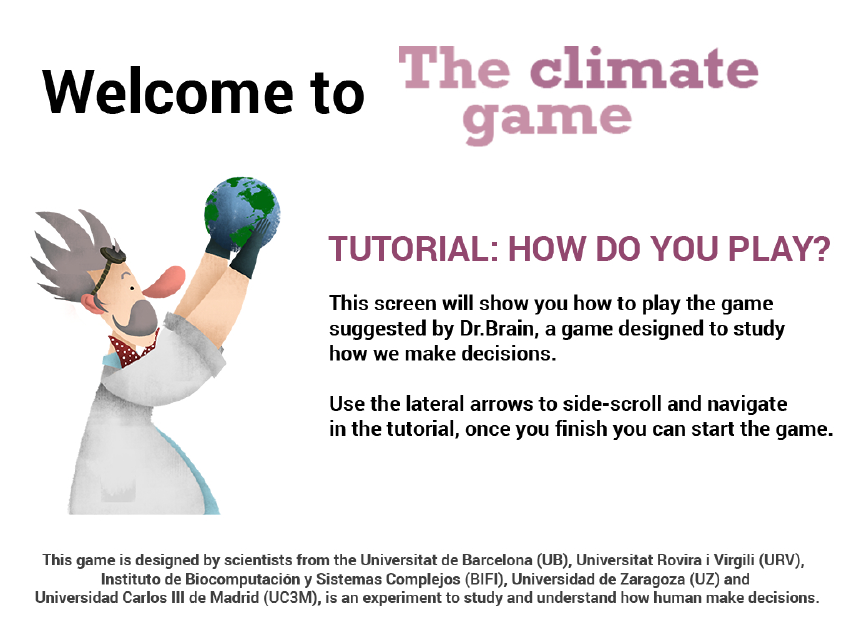}

\vspace{2mm} 

(b)
\includegraphics[width=0.3\textwidth]{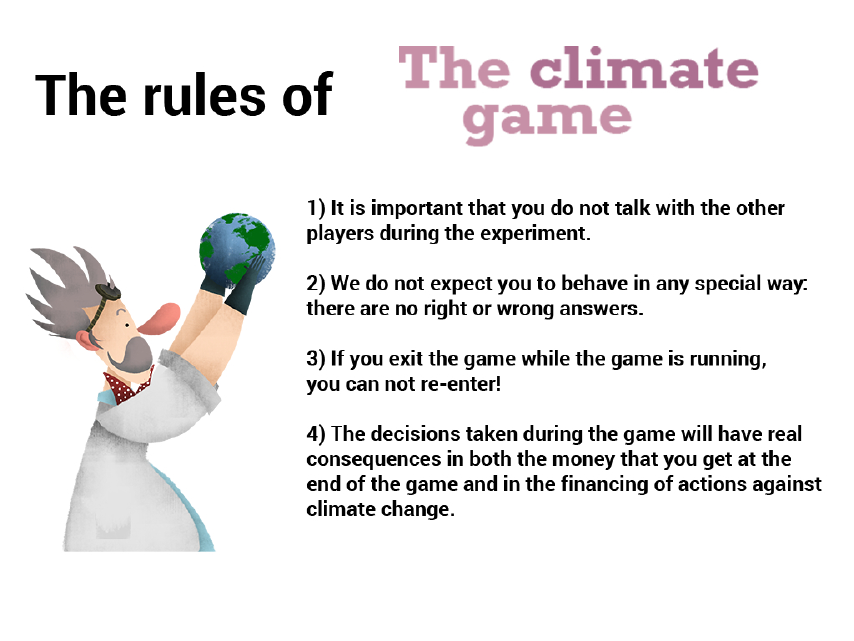}

\vspace{2mm}

(c) 
\includegraphics[width=0.3\textwidth]{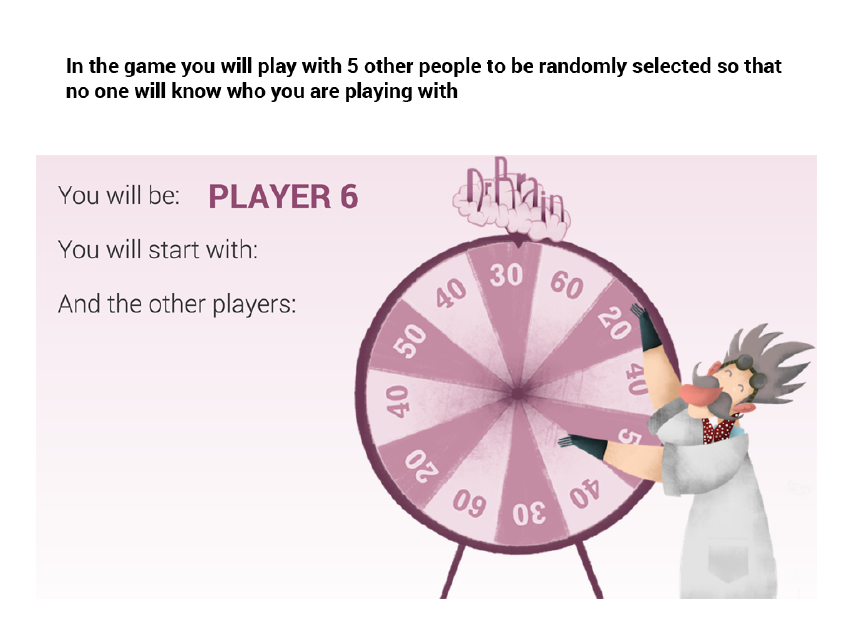}\\

\vspace{2mm} 

(d)
\includegraphics[width=0.3\textwidth]{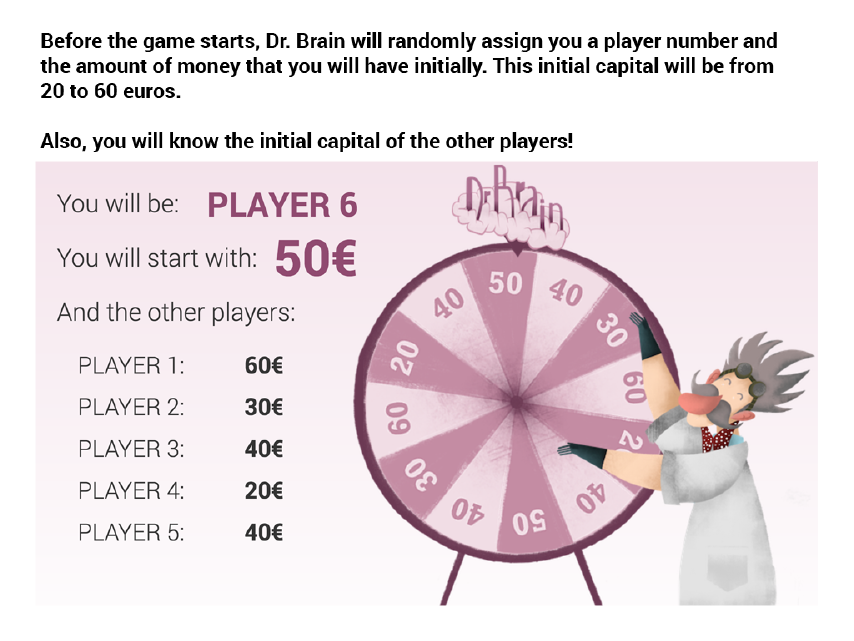}

\vspace{2mm} 

(e)
\includegraphics[width=0.3\textwidth]{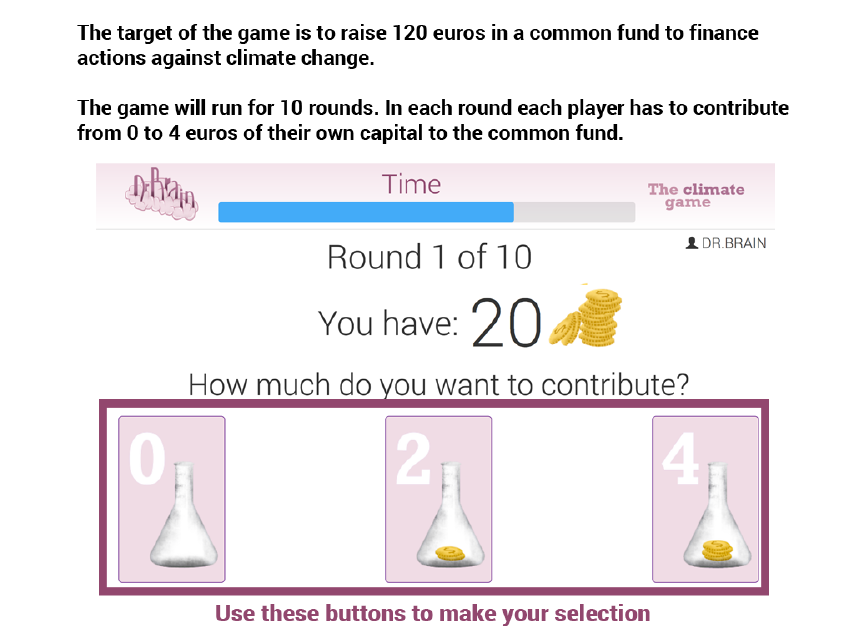} 

\vspace{2mm} 

(f)
\includegraphics[width=0.3\textwidth]{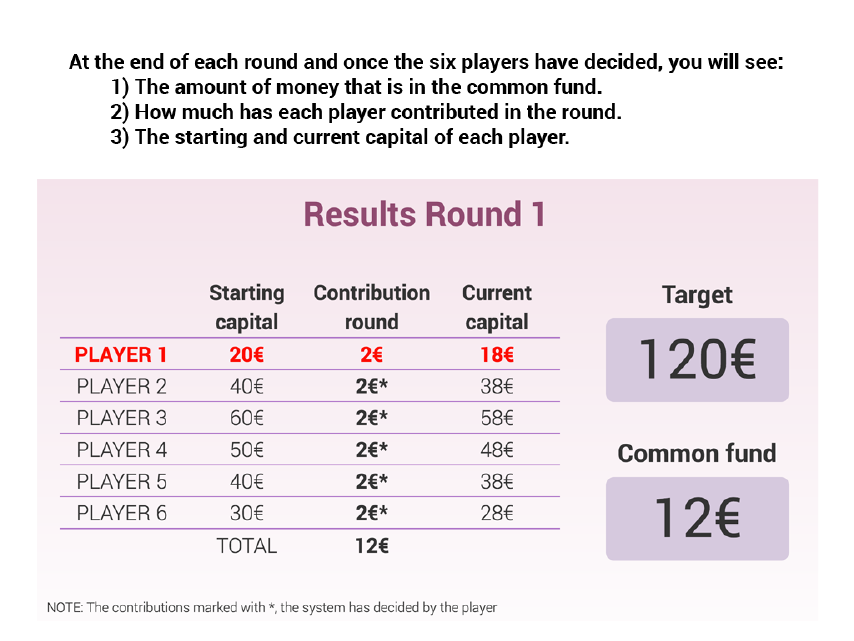}\\

\vspace{2mm} 

(g)
\includegraphics[width=0.3\textwidth]{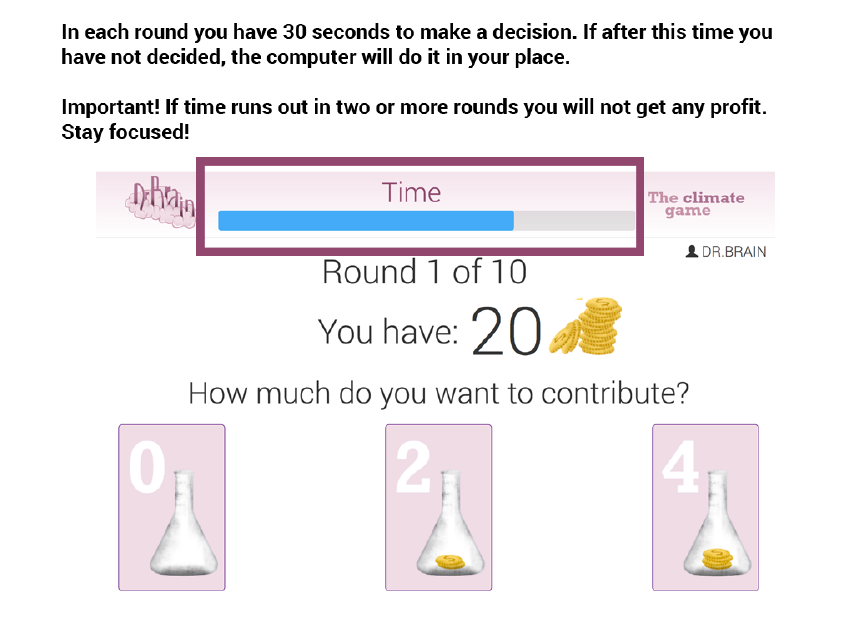} 

\vspace{2mm} 

(h)
\includegraphics[width=0.3\textwidth]{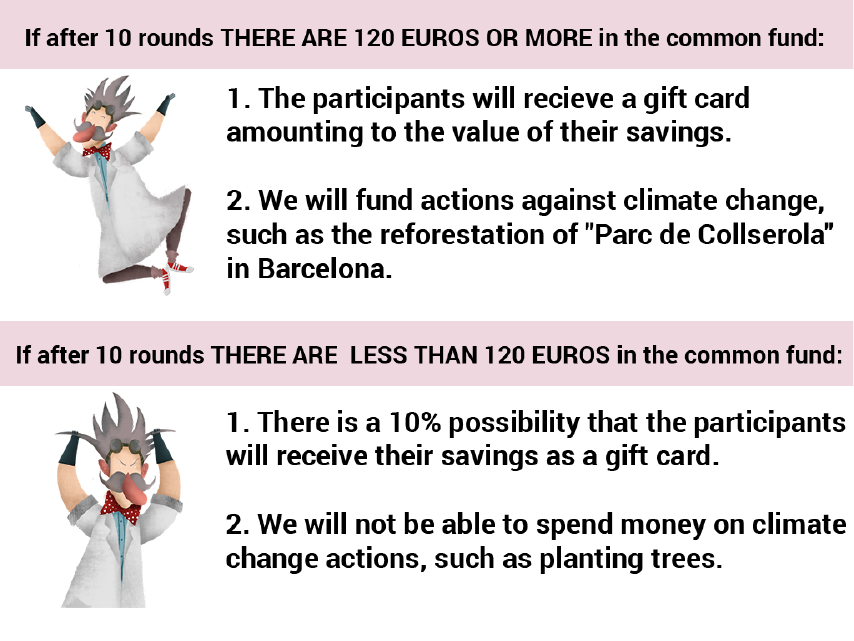}

\vspace{2mm} 

(i)
\includegraphics[width=0.3\textwidth]{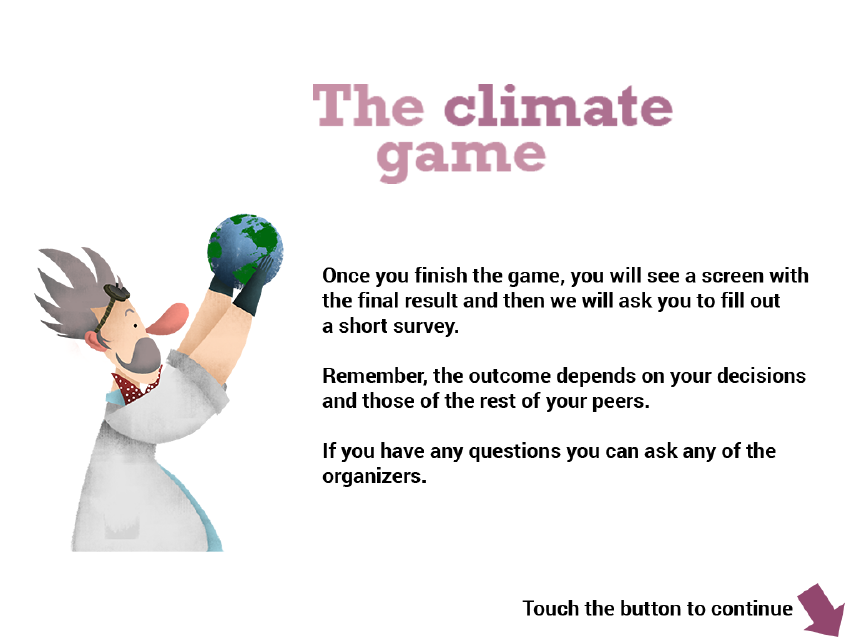} 

\end{array}$
\end{center}
\caption{Screen shots of the tutorial shown before the experiment.}
\label{fig:tutorial}
\end{figure*}

Tutorial Screen 1 (Fig. \ref{fig:tutorial}(a)). \textit{Welcome to The Climate Game. TUTORIAL: HOW DO YOU PLAY?. This screen will show you how to play the game suggested by Dr.Brain, a game designed to study how we make decisions. This game is designed by scientists from the Universitat de Barcelona (UB), Universitat Rovira i Virgili (URV), Instituto de Biocomputación y Sistemas Complejos (BIFI), Universidad de Zaragoza (UZ) and Universidad Carlos III de Madrid (UC3M), is an experiment to study and understand how human make decisions. Use the lateral arrows to side-scroll and navigate in the tutorial, once you finish you can start the game.}

Tutorial Screen 2 (Fig. \ref{fig:tutorial}(b)). \textit{The rules of The Climate Game. 1) It is important that you do not talk with the other players during the experiment. 2) We do not expect you to behave in any special way: there are no right or wrong answers. 3) If you exit the game while the game is running, you can not re-enter! 4) The decisions taken during the game will have real consequences in both the money that you get at the end of the game and in the financing of actions against climate change.}

Tutorial Screen 3 (Fig. \ref{fig:tutorial}(c)). \textit{In the game you will play with 5 other people to be randomly selected so that no one will know who you are playing with}

Tutorial Screen 4 (Fig. \ref{fig:tutorial}(d)). \textit{Before the game starts, Dr. Brain will randomly assign you a player number and the amount of money that you will have initially. This initial capital will be from 20 to 60 euros. Also, you will know the initial capital of the other players!}

Tutorial Screen 5 (Fig. \ref{fig:tutorial}(e)).\textit{The target of the game is to raise 120 euros in a common fund to finance  actions against climate change. The game will run for 10 rounds. In each round each player has to contribute from 0 to 4 euros of their own capital to the common fund.}

Tutorial Screen 6 (Fig. \ref{fig:tutorial}(f)).\textit{At the end of each round and once the six players have decided, you will see: 1) The amount of money that is in the common fund. 2) How much has each player contributed in the round. 3) The starting and current capital of each player.}

Tutorial Screen 7 (Fig. \ref{fig:tutorial}(g)).\textit{In each round you have 30 seconds to make a decision. If after this time you have not decided, the computer will do it in your place. Important! If time runs out in two or more rounds you will not get any profit. Stay focused!}

Tutorial Screen 8 (Fig. \ref{fig:tutorial}(h)).\textit{If after 10 rounds THERE ARE 120 EUROS OR MORE in the common fund: 1. The participants will recieve a gift card amounting to the value of their savings. 2. We will fund actions against climate change, such as the reforestation of "Parc de Collserola" in Barcelona. If after 10 rounds THERE ARE  LESS THAN 120 EUROS in the common fund: 1. There is a 10\% possibility that the participants will receive their savings as a gift card. 2. We will not be able to spend money on climate change actions, such as planting trees.}

Tutorial Screen 9 (Fig. \ref{fig:tutorial}(i)).\textit{The Climate Game. Once you finish the game, you will see a screen with the final result and then we will ask you to fill out a short survey. Remember, the outcome depends on your decisions and those of the rest of your peers. If you have any questions you can ask any of the organizers. Touch the button to continue}

\section*{Supplementary Figures}

\end{document}